\documentclass[a4paper,11pt]{article}
\usepackage[utf8]{inputenc}
\usepackage{natbib}
\usepackage{appendix}
\usepackage{upgreek}
\usepackage{amsmath}%
\usepackage{amsfonts}%
\usepackage{amssymb}%
\usepackage{amsthm}%
\usepackage{graphicx}
\usepackage{xcolor}
\usepackage{multicol}
\usepackage{enumerate}
\usepackage[lofdepth,lotdepth]{subfig}
\usepackage{authblk}
\usepackage{epstopdf}
\usepackage{epsfig}
\usepackage{float}
\usepackage{amsmath}
\usepackage{amssymb}
\usepackage{array}
\usepackage{latexsym}

\newtheorem{corollary}{Corollary}
\newtheorem{proposition}{Proposition}
\newtheorem{definition}{Definition}
\newtheorem{lemma}{Lemma}

\begin{document}
\title{Economic development and the structure of cross-technology interactions\footnote{This research is part of the Swiss Competence Center for Energy Research SCCER CREST of the Swiss Innovation Agency Innosuisse and has been supported by the Swiss National Science Foundation under Grant 100018\_169764.}\hspace{0.2cm}\footnote{\copyright 2021. This manuscript version is made available under the CC-BY-NC-ND 4.0 license http://creativecommons.org/licenses/by-nc-nd/4.0/}}
\author{Anton Bondarev\thanks{Corresponding author, International Business School Suzhou,  Xi'an Jiaotong-Liverpool University, Chongwen Road, 8, 215123 Suzhou, P. R. China, e-mail: anton.bondarev@xjtlu.edu.cn}
\and Frank C. Krysiak\thanks{Department of Business and Economics, University of Basel, Peter Merian-Weg 6, 4002 Basel, Switzerland, e-mail: frank.krysiak@unibas.ch}}
%\affil[1]{Internationakl}
%\affil[2]{University of Basel, Department of Business and Economics}

\date{December 2020}

\maketitle

{\renewcommand{\baselinestretch}{1.0}

\begin{abstract}
Most explanations of economic growth are based on knowledge spillovers, where the development of some technologies facilitates the enhancement of others. Empirical studies show that these spillovers can have a heterogeneous and rather complex structure. But, so far,  little attention has been paid to the consequences of different structures of such cross-technology interactions: Is economic development more easily fostered by homogenous or heterogeneous interactions, by  uni- or bidirectional spillovers? Using a detailed description of an r\&d sector with cross-technology interactions embedded in a simple growth model, we analyse how the structure of spillovers influences growth prospects and growth patterns. We show that some type of interactions (e.g., one-way interactions) cannot induce exponential growth, whereas other structures can. Furthermore, depending on the structure of interactions, all or only some technologies will contribute to growth in the long run. Finally, some spillover structures can lead to complex growth patterns, such as technology transitions, where, over time, different technology  clusters are the main engine of growth.
\end{abstract}}

% 168 words in abstract


\thispagestyle{empty}

\vfill

\pagebreak

\setcounter{page}{1}

\section{Introduction}
Technological progress is considered to be one of the most important drivers of economic growth. In fact, this idea is one of the foundations of modern endogenous growth theories, see \cite{Romer}, \cite{AH}, \cite{jones1999growth}, \cite{galor2000population}, and \cite{acemoglu2009introduction} among many others. 

Economic models have for a long time emphasized the importance of r\&d spillovers between technologies for technological change, see, for example, \cite{peretosmulders2002}. Many modern growth models rely on knowledge spillovers as the main driver of technological progress and thus long-term growth, see \cite{sorger2010}, \cite{Lucas2014}, \cite{acemoglu2015}, \cite{Chu2017}, \cite{hamano2017}, and \cite{Peretto2018} for recent examples. It is well established that the magnitude of such spillovers is an important determinant of growth. For example, in \cite{Bresnahan1995}, growth is traced to the unidirectional global spillovers coming from general purpose technologies (GPTs). 

However, recent empirical evidence suggests that spillovers are unevenly distributed across different technologies. \cite{Acemoglu2016}  find, using  US panel data, that the structure of spillovers is highly heterogeneous:  There are several central sectors/technologies that produce many unidirectional spillovers to other sectors, whereas the majority of sectors do not produce spillovers to other sectors but enjoy intra-sectoral spillovers. 

This indicates that  the structure of spillovers might warrant a closer investigation. But so far, there are few studies that analyse how a heterogeneous structure, as the one found by  \cite{Acemoglu2016}, influences overall knowledge spillovers' intensity and growth. For example, does a spillover structure where each technology contributes to and benefits from a common stock of knowledge lead to qualitatively similar implications as a structure where each technology only influences a small set of other technologies? Or does an economy with unidirectional r\&d spillovers experience similar growth patterns as one where spillovers are bidirectional? Is heterogeneity of  spillovers conducive or detrimental to overall growth?

\cite{peretosmulders2002} analyse an important point in this context. They use a setting where spillovers occur only among firms that are part of the same spillovers-network (i.e., whose technological developments  share a common core idea) and where firms developing new technologies may create new spillovers-networks (i.e., create a new core idea). They show that the scale effect vanishes asymptotically, as more and more networks are created, so that the size of each individual network (and thus the amount of spillovers received by an individual firm) will eventually cease to increase. 

Some other studies have also adopted a network structure of cross-sectoral spillovers, see, for example, \cite{Acemoglu2012} and \cite{Acemoglu2016}, while another strand of literature focusses on the difference between internal and external innovations and the resulting heterogeneity in qualities \citep{Akcigit2018}. The studies on spillover networks have shown that network structures have important consequences as to how shocks are transmitted, attenuated or amplified in an economy (see, e.g., \cite{Arata2017}, \cite{Enghin18}, \cite{Gualdi18}, \cite{Oberfield18}), whereas the second strand of literature provides important insights into the effects of firm sizes and market positions on innovation and thus on technological changes.

However, a general assessment of how different structures of spillovers influence growth patterns and growth prospects is missing, although this question is of substantial importance: Different structures of spillovers could lead to different growth prospects. For example, spillovers might induce exponential or only linear growth. They could also  induce different patterns of growth, for example, all technologies could contribute to growth, only a few technologies could drive growth in the long run, or there could be transitions between different clusters of technologies that each are the main driver of economic growth for some time. 

Indeed, there  is some empirical evidence of a slowdown in economic growth and in r\&d productivity \citep{gordon2016,jones2017}, so that it might be important to explain development patterns  apart from exponential growth. A more detailed study of spillover structures could offer one option to this end. 

In this paper, we investigate how the structure of spillovers between technologies influences  technological development in a decentralized economy and what consequences this implies for growth prospects and growth patterns. 

To this end, we use a highly stylistic growth model. As in \cite{acemoglu2010}, we consider an economy with a capital- and a labour-intensive (unskilled labour) sector, but, in our model, r\&d enhances only the productivity of labour. R\&D results from the efforts of r\&d firms that each develop a particular technology by employing scientists (skilled labour).

The novel aspect of our model is a detailed description of  spillovers between different technologies in r\&d. In our model of the r\&d sector,  each firm's r\&d decisions (and the resulting  technology's development which we refer to as quality development) are modeled as an optimal control problem. The firms' problems are coupled via a matrix that describes cross-technology interactions, such as knowledge spillovers. By using different types of matrices, we can model different structures of interactions, such as a common stock of knowledge (homogeneous spillovers), uni- or bilateral interactions, and (circular) chains of technological development.

Our results indicate that, in our model, the \emph{structure} of spillovers is indeed an important determinant of long-run growth prospects and patterns.  We show that an economy has to include either intra-technology spillovers or at least one circular chain of cross-technology interactions to exhibit  r\&d-driven exponential growth. Whereas the often used model of a common pool of knowledge complies with this assumption, many other structures of r\&d interactions do not.

Moreover, our results show that only technologies that receive direct or indirect spillovers  from the fastest growing technology will be developed in the long run. Also, there can be a particular form of a path-dependency: Initial conditions (past development efforts) determine which technology cluster will be the one that drives growth in the long run. This can, for example, occur if there are different clusters of technologies that are separated from each other (i.e., there are no spillovers between these clusters).
 
 A particularly interesting case is that of one-way spillovers between different groups of technologies, where a cluster of technologies provides a spillover to another cluster but not vice versa. In such a setting, technology transitions can be described: Growth will be first driven by one and later by the other cluster. This can cause rather complex growth patterns.

Compared to \cite{peretosmulders2002}, we do not cover an expanding variety of technologies but allow for more general spillover structures. Whereas \cite{peretosmulders2002} assume that spillovers exist only within a network and are homogeneous wherever they exist, we allow for arbitrary linear spillover structures. In particular, our spillovers can be asymmetric, which is impossible even in the most general version of \cite{peretosmulders2002}. Several interesting growth patterns found in our analysis are linked to heterogeneous or even asymmetric spillovers, such as the above-mentioned technology transitions. Consequently, our analysis should be seen as a complement to the study of \cite{peretosmulders2002}: We analyze the effects of different structures of spillovers in more detail but abstract from the variety expansion that is central to their results.

In the next section, we  set up the model. In Section \ref{sec_sol}, we study the dynamics of technological change and show how growth prospects and growth patterns depend on the structure of knowledge spillovers. Section \ref{sec:ext} extends our results and discusses the relevance of some model assumptions. Section \ref{sec.concl} concludes.

\section{A growth model with cross-technology interactions}
\label{sec.mod}

We use a model that adapts the setting of \cite{acemoglu2010} to a situation with cross-technology spillovers. In \cite{acemoglu2010}, there are $n$ labour-intensive and $m$ capital-intensive intermediate goods that are used to produce a labour- and a capital-intensive aggregate good. R\&D can increase the number of varieties of these intermediate goods and is based (in the specification used in most parts of the paper) on a fixed amount of skilled labour (scientists). Furthermore, \cite{acemoglu2010} shows that along a BGP, r\&d will increase the variety of labour-intensive goods, whereas an increasing capital stock will facilitate growth in the capital-intensive part of the economy.

We adjust this setting to accommodate spillovers between different technologies by assuming that the number of intermediate goods is constant, with $N$ labour-intensive intermediates and a single capital-intensive intermediate, but that each labour-intensive good is produced with a technology whose quality (i.e., the marginal productivity of labour) can be enhanced by r\&d. This r\&d is conducted by a constant number of scientists that are employed by r\&d firms. The productivity of the scientists in advancing a given technology depends on spillovers from other technologies. In contrast, the capital-intensive sector can grow solely via capital accumulation. 

In detail, consider an economy with a standard representative consumer, having  the utility function:
\begin{align}
\max_{C\geq 0}\;\int_{0}^{\infty}\mathrm{e}^{-\rho\:t}U(C)dt.
\end{align}
where $C$ is consumption of a final good. This good  can also be used as an investment good in the capital-intensive sector, with investment being denoted by $I$. The consumer supplies one unit of unskilled labour (wage: $w$) as well as $S$ units of skilled labour used in r\&d (scientists, wage $w_S$).

The budget constraint of the consumer can be written as
\begin{align}\label{budc}
C+I\leq Y,
\end{align}
where $Y=\left(Y^{(\epsilon-1)/\epsilon}_{L}+Y^{(\epsilon-1)/\epsilon}_{K}\right)^{\frac{\epsilon}{\epsilon-1}}$ denotes the aggregate of the labour- and capital intensive goods $Y_L$ and $Y_K$, respectively (as is common, we assume $\epsilon>1$). We use the price of $Y_L$ as numeraire.

The capital-intensive good is produced by a price-taking firm with a technology $Y_K=\gamma \: K$, where $\gamma>0$ and $K$ is capital that is accumulated according to
\begin{align}\label{eq.Kdot}
	\dot{K}=I.
\end{align}
The labour-intensive good is also produced by a price-taking firm but from a set of $N$ labour-intensive intermediates via the technology 
\begin{align}\label{eq.yl}
Y_L=\left(\sum_{i=1}^{N}y_i^\beta\right)^{1/\beta},
\end{align}
with $\beta\in]0,1[$. To render the r\&d model tractable, we set $\beta=1/2$. This assumption helps us to focus our analysis on the influence of spillovers, as it renders the incentive to increase quality homogeneous among firms, so that r\&d firms solely differ with regard to the spillovers they cause or receive (see Sect. 3). As we will argue later on (see Sect. 4), this assumption has only a limited influence on our qualitative results but it facilitates a closed-form solution of the dynamic optimization problem of the r\&d firms.

Each of the labour-intensive intermediates is produced according to a technology
\begin{align}\label{eq.yi}
y_i=q_i(t)\:l_i,
\end{align}
where $l_i$ is labour used for intermediate $i$ and $q_i(t)$ is the quality of the technology at time $t$ that is used to produce this intermediate. The firms producing these intermediates act as monopolists on their product market but face perfect competition on the labour market.

Due to Eq. \eqref{eq.yl}, the demand for the good of monopolist $i$ is isoelastic, so that (together with $\beta=1/2$), we can write the inverse demand function in the form
\begin{align}\label{eq.pi}
p_i=c\:y_i^{-1/2}, 
\end{align}
where $c$ is a factor that is identical for all monopolists.

Solving the monopolists' optimization problem yields $l_i=\frac{c^2}{4\: w^2}\:q_i(t)$, which implies $p_i=2 w/q_i$. Note that this is in line with the usual result that the price of intermediate $i$ is simply a markup on the marginal labour costs, which in our setting are $w/q_i(t)$ (as the firm needs $1/q_i(t)$ units of labour to produce one unit of output).

This labour supply implies that monopolist $i$ achieves a profit
\begin{align}\label{eq.profiti}
\tilde{\pi}_i=p_i\:q_i(t)\:l_i-w\:l_i=\frac{c^2}{4\: w}\: q_i(t). 
\end{align}
From our above analysis, it follows that the wage has to be proportional to $\sqrt{\sum_{j=1}^N q_j(t)}$, so that demand and supply for unskilled labour match. 

Finally, our specification implies that the consumer's optimal consumption path is characterized by the standard Euler equation:
 \begin{align}
 g_{C}=\frac{\dot{C}}{C}=(r(t)-\rho), 
 \end{align}
which together with the constant marginal productivity of capital defines the interest rate and the accumulation of capital.
 
 For our further analysis of r\&d-driven growth, the growth rate of $Y_L$ will be of particular interest, as the labour-intensive sector is the sector that derives its growth prospects purely from technological progress. Let $g_{Y_L}(t)$ denote this growth rate at time $t$ and let $g_{q_{i}}(t)$ denote the growth rate of the quality $q_i(t)$ of technology $i$ at time $t$. Then, our above analysis implies that
	\begin{equation}\label{eq.ygro}
	g_{Y_L}(t)=\sum_{i}^{N}g_{q_{i}}(t)\frac{q_{i}(t)}{\sum_{j}^{N}q_{j}(t)}.
	\end{equation}
	This relation has two important properties:
	\begin{enumerate}
		\item If the growth rates of all qualities are equal ($g_{q_{i}}(t)=g_q$), then the growth rate in the labour-intensive sector equals this growth rate as well ($g_{Y_L}(t)=g_q$).
		\item If, after some initial time $T$, the qualities of some technologies grow at the same rate ($g_{q_{i}}(t)=g_q$ $\forall t>T$), whereas others do not grow at all ($g_{q_{j}}(t)=0$ $\forall t>T$), then the growth rate in the labour-intensive sector will converge to the growth rate of the growing technologies.
	\end{enumerate}
Thus, in two cases that will be of particular importance for our analysis, we get the simple result that growth in the labour-intensive sector is simply proportional to quality growth.
	
In our subsequent analysis, we will only focus on growth in the labour-intensive sector, as this is the r\&d-driven part of growth and r\&d is the focal point of this paper. Furthermore, due to Eq.  \eqref{eq.ygro}, we will usually discuss growth in terms of quality growth, as the growth rate of the labour-intensive sector is simply a weighted average of the growth rates of the qualities of the different technologies.

\subsection{The r\&d sector}\label{sec:rd}
For the r\&d sector, we assume that each technology $i$ ($i\in{1,\ldots,N}$) is developed and marketed by a single firm. All r\&d firms use skilled labour (scientists, total quantity $S$) as their sole input for technology development. Thus, the r\&d process is described as a knowledge-based one, following the definition of \cite{rivera1991} and others.  The state of each technology $i$ at time $t$ is characterized by its quality $q_{i}(t)$. 

For simplicity, we assume that the number of sectors/technologies is constant over time, that is, there are no horizontal innovations. 

Each r\&d firm is a monopolist in supplying technology $i$. But there is perfect competition on the market for scientists. There are no new versions of each technology. Instead, each firm is continuously developing its technology based on profit incentives coming from the payments of  the firms using this technology for producing an intermediate good $y_{i}$. The r\&d sector is thus described by the usual monopolistic competition as in many recent growth papers (see, e.g., \cite{Akcigit2018}, \cite{myMD}, \cite{Peretto2018}). As each r\&d firm is a monopolistic supplier of its technology and as all firms using this technology are homogeneous, the price for this technology (e.g., a patent payment) will be set at each point of time (that is, for each set of technological qualities) to absorb all profits made in intermediate goods production.

Given this, each r\&d firm maximizes its discounted stream of profits:
\begin{align}
V_{i}=\int_{0}^{\infty}\mathrm{e}^{-r(t)\:t}\left(\Pi_{i}(q_i(t),t)-\omega_S(t)\:S_i(t)\right)dt,
\end{align}
where $\Pi_{i}(q_i(t))$ is the profit flow from selling the technology $i$ at quality $q_i(t)$ to the associated firm in the manufacturing sector, where $\omega_S(t)$ is the wage of scientists, and where $S_i(t)$ is the number of scientists hired by r\&d firm $i$ to enhance its technology.

The last necessary ingredient is the production of knowledge for each sector. To this end, we use a specification similar to \cite{acemoglu2010}, but use a knowledge spillover between the different technologies:\footnote{\cite{acemoglu2010} uses the more general specification (in our notation and adjusted to the multi-technology setting) $g^{1-\eta}_{i}\:\left(S^{\nu}_{i}\:\left(F_{i}\:\mathbf{q}+\alpha\right)\right)^{\eta}$ but sets $\eta=1$ for most parts of the analysis, which results in a model close to our specification for the case of a single technology.}
\begin{align}\label{eq.rd}
\dot{q}_{i}=S^{\nu}_{i}\:\left(F_{i}\:\mathbf{q}+\alpha\right),
\end{align}
where we assume $0<\nu<1$.

In the above equation, $F_{i}\:\mathbf{q}$ are the spillovers coming from all other firms to the firm $i$ as a function of the vector of all qualities $\mathbf{q}$. If we consider all firms, the vector $F_{i}$ that describes the spillovers to firm $i$ can be seen as the $i$-th row vector of a spillover matrix $\mathbf{F}$ that contains all spillovers to and from all firms. We focus our attention on positive effects between technologies and thus assume that all elements $F_{ij}$ of $\mathbf{F}$ are non-negative.

It is instructive to compare our setting to that of \cite{peretosmulders2002}. There, spillovers exist only within networks and are symmetric and of identical intensity wherever they exist. The case where each firm is part of exactly one spillovers-network (which is the main case analysed in \cite{peretosmulders2002}), can thus be depicted in our setting as a block-diagonal matrix $\mathbf{F}$  whose non-zero entries all have the same value.\footnote{The case, where firms select the networks they would like to belong to, cannot be replicated in our setting, as we take the spillover structure as being given and not subject to choice.} In contrast, we assume a constant number of technologies, whereas the central element of \cite{peretosmulders2002} is an expanding variety of technologies that induces a dilution of public knowledge. Given these points, our setting should be seen as a complement to \cite{peretosmulders2002}, as we allow for a variety of spillover structures but use less sophisticated assumptions on the set of available technologies.

The dynamics of \eqref{eq.rd} are our main interest in this paper. We will analyse the dependency of these dynamics on the properties of the spillover matrix $\mathbf{F}$. The r\&d firm decides about the optimal allocation of research labour but does not take into account decisions of other firms affecting $\mathbf{q}$, treating $F_{i}\:\mathbf{q}$ as an exogenous function of time. Consequently, r\&d is described in our model as a system of $N$ coupled dynamic programming problems.

\section{The structure of cross-technology interactions and long-run development}\label{sec_sol}
To analyse the impact of different structures of cross-technology interactions on the prospects for long-run growth, we analyse our model for different spillover structures, that is, for different types of matrices $\mathbf{F}$.  

As a first step, we  characterize the optimal behaviour of the r\&d firms.
Based on the model introduced above, the intertemporal optimization problem of firm $i$ can be written as
\begin{align}
\max_{g_{i}(t)\geq0}&\int_{0}^{\infty}\mathrm{e}^{-r(t) \: t}\left(\Pi_{i}(q_i(t),t)-\omega_S(t)\:S_i(t)\right)dt,\label{obji}\\\nonumber &\text{s.t.}\\
\dot{q}_{i}&=S^{\nu}_{i}(t)\:\left(F_{i}\:\mathbf{q}(t)+\alpha\right).\label{qtoti}
\end{align}
As each firm $i$ accounts only for the second effect in Eq. \eqref{qtoti} (the $\alpha_i$), the term $F_i\:\mathbf{q}(t))$ in Eq. \eqref{qtoti} is, from this firm's perspective, simply  some function of time $f_i (t)$. Taking into account Eq. \eqref{eq.profiti}, the decision problem of firm $i$ is characterized by the following Hamiltonian:
\begin{equation}\label{eq.Hi}
	H_i=\gamma(t)\:q_i(t)-\omega_S(t)\:S_i(t)+\psi_i(t)\:S^{\nu}_{i}(t)\:\left(f_i(t)+\alpha\right),
\end{equation}
where $\gamma(t)$ is the marginal return to quality, which is identical for all firms according to Eq. \eqref{eq.profiti}, and where $\psi_i(t)$ is the co-state variable of firm $i$'s decision problem.

Calculating the firm's optimal demand for scientists under the assumption that $F_i\:\mathbf{q}(t)\geq 0$, which is met due to all entries of $F_i$ as well as all $q_i(t)$ being non-negative,  leads to 
\begin{equation}
	S_i(t)=\left(\frac{\nu\:\psi_i(t)\:(f_i(t)+\alpha)}{\omega_S}\right)^{\frac{1}{1-\nu}}.
\end{equation}
Market clearing on the market for scientists implies $\sum_{j=1}^N S_j (t)=S$ at all times, which yields	
\begin{equation}
	\omega_S^{\frac{1}{1-\nu}}(t)=\frac{\left(\nu\:\sum_{j=1}^N\psi_j(t)\:(f_j(t)+\alpha)\right)^{\frac{1}{1-\nu}}}{S},
\end{equation}
and thus
\begin{align}
	S_i(t)&=s_i(t)\: S, \\
	s_i(t)&=\frac{\left(\psi_i(t)\:(F_{i}\:\mathbf{q}(t)+\alpha)\right)^{\frac{1}{1-\nu}}}{\sum_{j=1}^N \left(\psi_j(t)\:(F_{j}\:\mathbf{q}(t)+\alpha)\right)^{\frac{1}{1-\nu}}}.
	\end{align}
In this specification, $s_i(t)$ is simply the share of scientists employed by r\&d firm $i$ at time $t$. To gain insight into these shares, we have to assess the co-state variables. By Eq. \eqref{eq.Hi}, these follow from
\begin{equation}\label{dynpsi}
	\dot{\psi}_{i}(t)=r(t) \:\psi_{i}(t)-\gamma(t).
\end{equation}
As this equation is independent of $q_i(t)$, there are no firm-specific parameters, and all r\&d firms have identical transversality conditions, the co-states of all firms will be identical at all points of time.\footnote{This is where the assumption $\beta=1/2$ is essential; only due to this assumption the co-state equation is independent of qualities, which allows to solve the model. Economically, this implies that increasing a quality by one unit yields the same payoff for all technologies at all times, which explains why the co-state variables take on identical values for all firms.}  This implies that we can write the share of scientists employed by r\&d firm $i$ in the simplified form
\begin{equation}\label{shares.1}
	s_i(t)=\frac{\left(F_{i}\:\mathbf{q}(t)+\alpha\right)^{\frac{1}{1-\nu}}}{\sum_{j=1}^N \left(F_{j}\:\mathbf{q}(t)+\alpha\right)^{\frac{1}{1-\nu}}}.
\end{equation}
This equation shows that the share of scientists that an r\&d firm employs at time $t$ depends on the spillovers that this firm receives relative to the spillovers other firms get. In fact, the above equation implies
\begin{equation}\label{shares.2}
	\frac{s_i(t)}{s_j(t)}=\left(\frac{F_{i}\:\mathbf{q}(t)+\alpha}{F_{j}\:\mathbf{q}(t)+\alpha}\right)^{\frac{1}{1-\nu}}.
\end{equation}
This is, of course, a consequence of the competition for scientists on the skilled labour market; firms that receive higher spillovers have a higher marginal productivity of scientists and will thus employ more scientists at the same equilibrium wage.

Note that Eq. \eqref{shares.1} implies that the allocation of scientists will converge  to some constant allocation over time, if there is at least a single technology that has unlimited growth and causes a spillover. Furthermore, technologies that do not receive spillovers at all (or that only receive spillovers from technologies whose quality does not grow rapidly enough) will have a declining share of scientists that converges to zero in the long run.

The following first result formalizes these insights to a characterisation of the long-run distribution of scientists among r\&d firms.
 \begin{lemma}\label{lem:1}
Let $\mathcal{I}=\{1,\ldots,N\}$ and assume that there is at least one technology with unbounded growth that causes a spillover, that is, there are $i,j\in\mathcal{I}$ with $\lim_{t\rightarrow\infty}q_i(t)\rightarrow\infty$ and  $F_{ij}>0$.

Then, the set $\mathcal{I}$ (all technologies) can be split into two disjoint sets $\mathcal{I}_0$ (technologies that stop to grow, can be the empty set) and  $\mathcal{I}_\infty$ (technologies that continue to grow) and the distribution of scientists among r\&d firms converges to the following steady state: 
\begin{equation}\label{shares.2b}
	s_i^{\infty}=\begin{cases}
	0,&\text{ for }\:\: i\in\mathcal{I}_0,	\\
	\frac{\left(\sum_{j\in\mathcal{I}_\infty}F_{ij}\:\zeta_{j}\right)^{\frac{1}{1-\nu}}}{\sum_{j\in\mathcal{I}_\infty}\left(\sum_{k\in\mathcal{I}_\infty}F_{jk}\:\zeta_{k}\right)^{\frac{1}{1-\nu}}},&\text{ for }\:\: i\in\mathcal{I}_\infty,
\end{cases}
\end{equation}
where $\zeta_{i}\in[0,1]$ for all $i\in\mathcal{I}_\infty$. Furthermore, all technologies in the set $\mathcal{I}_\infty$ (technologies that continue to grow) will have identical growth rates in the long run.
\end{lemma}
\begin{proof} See Section \ref{proof1} in the appendix.\end{proof}
The lemma shows that, in the long run, only those technologies will continue to develop that are connected directly or indirectly to the most rapidly growing technology. This is a result of the competition for scientists. Technologies that receive spillovers from the most rapidly growing technologies have a higher marginal productivity of scientists. In the long run, this difference becomes so large that these technologies will attract all available scientists, despite the assumed declining marginal productivity of scientists ($\nu<1$) in Eq. \eqref{eq.rd}.

That spillovers largely determine which technologies are going to be developed in the long run is a consequence of the homogeneity of the different technologies in the production sector: In our model, a marginal increase of the quality of a technology leads to the same increase in profit for all technologies, as Eq. \eqref{eq.yl} is symmetric with regard to technologies and as our assumption of $\beta=1/2$ implies that the marginal benefit of technology development is constant (i.e., not changing with the quality). Due to this, the incentive to increase quality is the same for all technologies; they only differ with regard to how many scientists are required to achieve a given quality increase. This difference is determined by the spillovers and therefore the spillovers largely determine the long-run distribution of scientists and thus, via Eq. \eqref{qtoti}, the long-run dynamics.

The above result is central to our analysis of the long-run dynamics. It shows that, when investigating technological development in the long run, we can safely neglect reallocations of scientists among the r\&d firms that remain active. Such a reallocation will have substantial consequences in the short run (see Prop. 3 below), but after some time the allocation becomes stable and thus has no further consequences for growth patterns. This implies that the long-run dynamics in our model are governed by the following set of ODEs, whenever we have continued quality growth:
\begin{align}\label{eq.genrda}
\dot{\mathbf{q}}(t)=\mathbf{S}\:(\mathbf{F}\:\mathbf{q}+\mathbf{\alpha}),
\end{align}
where $\mathbf{S}$ denotes a diagonal matrix whose diagonal elements equal $(S_i^\infty)^\nu$, which are the long-run numbers of scientists employed by the different r\&d firms. Note that this matrix $\mathbf{S}$ alters the relative size of entries in $\mathbf{F}$ but does not affect its structure for the surviving technologies, as it is a diagonal matrix. The exception is, of course, that some firms will not be able to hire any scientists in the long run, implying that the respective entries in $\mathbf{S}$ are zero and, consequently, the corresponding qualities remain constant. Observe further that the above ODEs are linear in qualities, so that they allow for the option of quality growth at constant rates. 

For simplicity, we rewrite Eq. \eqref{eq.genrda} as
\begin{align}\label{eq.genrda2}
\dot{\mathbf{q}}^{\infty}(t)=\mathbf{F}^{\infty}\:\mathbf{q}^{\infty}+\mathbf{\alpha}^{\infty},
\end{align}
where
$\mathbf{q}^{\infty}$ is the vector of all technologies that are still developed in the long run, where 
$\mathbf{F}^{\infty}:= \mathbf{S}\:\mathbf{F}$ and where $\mathbf{\alpha}^{\infty}$ captures the constant part in Eq. \eqref{eq.genrda}, that is $\mathbf{S}\:\mathbf{\alpha}$, as well as the constant spillover from the technologies whose qualities have stagnated. 

Lemma \ref{lem:1} shows that some technologies can exhibit continued growth, whereas others might stagnate.  Which technologies will continue to grow is determined by the Eqs. \eqref{qtoti}, \eqref{shares.1}, and the initial conditions. This is a system of $N$ coupled nonlinear differential equations, which cannot be solved in general. However, it is possible to provide a simpler characterisation of the set of technologies that will grow in the long run.

To this end, we observe that, according to Lemma \ref{lem:1}, all technologies that have continued quality growth in the long run have to grow proportionally to each other, that is, they will have identical growth rates. Thus, we have to analyse whether the spillover structure combined with the market clearing allocation of scientists admits set(s) of technologies that grow at the same rate.

Define the ratio of a technology's quality to the average quality $Q(t)$ as $z_i(t):=q_i(t)/Q(t)$. As is obvious from this definition, all technologies $j$ that do not grow continuously will have a $z_j(t)$ that is zero in the long run. Furthermore, Lemma \ref{lem:1} implies that, for all technologies that continue to grow in the long run, the $z_i(t)$ will converge to a constant value. As all technologies that continue to grow have to grow at the same rate, we can characterize the set of technologies by the following system of equations and inequalities:
\begin{align}\label{eq.lr-1a}
&\forall i,j=1,\ldots, N, 	j>i: & z_i\:z_j\:(g_i(t)-g_j(t))=0,\\\label{eq.lr-1b}
&\forall i=1,\ldots, N:&z_i\geq0.
\end{align}
Here, $g_i(t)$ denotes the growth rate of technology $i$ and $\zeta_i$ is the long-run value of $z_i(t)$. Given this, Eq. \eqref{eq.lr-1a} simply states that, for every pair $i,j$ of technologies that continues to grow ($z_i,z_j>0$), the growth rates have to be identical.\footnote{If a technology $j$ stagnates, we have  $z_j=0$ and thus  all equations  in which this technology appears are always met.} Using our above definition of $z_i(t)$, with $\mathbf{z}$ being the vector of all $z_i$, as well as Eqs. \eqref{qtoti} and \eqref{shares.1}, we can write the above system for sufficiently large $t$ as 
\begin{align}\label{eq.lr-2a}
&\forall i,j=1,\ldots, N, 	j>i: & z_j\:\left(F_i\:\mathbf{z}\right)^{\frac{1}{1-\nu}}=z_i\:\left(F_j\:\mathbf{z}\right)^{\frac{1}{1-\nu}},\\\label{eq.lr-2b}
&\forall i=1,\ldots, N:&z_i\geq0,
\end{align}
whenever there is long-run growth. This leads to the following result.
\begin{proposition}\label{prop.lrgrowth}
	Assume that there is long-run growth, that is, there is at least one technology $i$ with $\lim_{t\rightarrow\infty} q_i(t)=\infty$, and that this technology causes a spillover. Then:
	\begin{enumerate}
		\item 	The system \eqref{eq.lr-2a}--\eqref{eq.lr-2b} has at least one solution $\mathbf{z}^{*}$ where at least one element of $\mathbf{z}^{*}$ is strictly positive. 
		\item Each solution $\mathbf{z}^{*}$ of Eqs. \eqref{eq.lr-2a}--\eqref{eq.lr-2b}  characterizes a candidate for the set of technologies that have continued growth: All technologies $i$ with $z_i^{*}>0$ have such growth, whereas all other technologies (with $z_j^{*}=0$) stagnate.
		\item If the system  \eqref{eq.lr-2a}--\eqref{eq.lr-2b} has a unique solution, then the set of technologies that continue to grow is fully determined by the spillover structure.
		\item If the system \eqref{eq.lr-2a}--\eqref{eq.lr-2b} has multiple solutions, then the set of technologies that continue to grow can depend on the spillover structure,  the value of $\alpha$, and the initial conditions.
	\end{enumerate}
\end{proposition}
\begin{proof} See Section \ref{proof1b} in the appendix.\end{proof}
This result reduces the question which technologies can potentially have continued quality growth  to the solution of an equation system (instead of analyzing a set of nonlinear differential equations). As this system is nonlinear, it can have a finite number of multiple solutions, which implies that there can be different technology clusters that are candidates for the long-run winner. If this is the case, the answer to the question which technology cluster will prevail in the long run  can depend both on the spillover structure and on initial conditions. 

We have already explained above that the spillover structure is clearly an important determinant for which technologies will grow in the long run. The insight provided by Prop. \ref{prop.lrgrowth}, that initial conditions, and thus past r\&d decisions, can also be important  is somewhat surprising. From an economic perspective, there are two main causes for this.

The first one is the interaction of spillovers and competition for scientists. At each point of time, a firm's ability to attract scientists depends on the spillovers that it receives and thus on the current distribution of quality among technologies; firms that receive spillovers from already fairly developed technologies have an advantage in the competition for scientists. But the allocation of scientists that results from this will determine the qualities in the next period. Now, if there are, for example, two clusters of technologies that each have strong spillovers within the cluster but no spillovers between clusters, only one of these clusters will see continued development according to Lemma \ref{lem:1}, as it is able to attract all scientists in the long run. Which of the two clusters will eventually claim this position depends on which cluster grows more rapidly and thus, potentially, on the initial shares of scientists that are strongly influenced by the initial qualities.

A second reason is that spillovers are externalities and we investigate the outcome of decentralized hiring decisions of scientists. Thus, while a firm takes into account the spillovers that it currently receives, when deciding how many scientists it hires at a given wage, it does not take into account that the result of this decision might influence the development of the technologies that create this spillover. This implies that the quality development is not perfectly coordinated by the market for scientists. From a societal perspective, scientists should be allocated both according to the spillovers that are received by a technology (which determine the current productivity of scientists) and the spillovers that are provided by a technology (which determine the societal benefit of developing the technology). But in the decentralized outcome analysed here, only the former effect is taken into account. This creates room for inefficient outcomes and thus for the particular type of path-dependency discussed above.

So far, our results all depend on the assumption that there is growth in the long run. As a next step, we investigate whether and how different types of spillovers induce growth. To this end, it is useful to distinguish different classes of spillovers. The matrix $\mathbf{F}$ that describes spillovers in our model is only a first step towards this end, as it describes solely the spillovers that occur directly between two technologies. But if a technology $i$ causes a spillover to a technology $j$, which in turn influences the development of a third technology $k$, then the technology $k$ receives an indirect spillover from $i$ via $j$. To capture these indirect effects,  we distinguish spillover structures by examining the network of spillovers, that is, by seeing the matrix $\mathbf{F}$ as the description of a directed graph. 

In the next steps of our analysis, only the existence of a spillover between two technologies (direct or indirect) matters, not the strength of the spillover. Therefore, we define an adjacency matrix that characterizes the structure of the network of spillovers. This adjacency matrix $\mathbf{A}$ is constructed from the spillover matrix $\mathbf{F}$ by setting the components $a_{i,j}$ of $\mathbf{A}$ in the following way
\begin{equation}\label{adjacency}
a_{i,j}=\begin{cases}
1,&\text{ if }\:\: f_{i,j}>0	\\
0,&\text{ otherwise.}
\end{cases}
\end{equation}
Thus, $a_{i,j}$ simply indicates whether technology $i$ receives a spillover from technology $j$ (in this case, $a_{i,j}=1$) or not ($a_{i,j}=0$). This matrix is particularly useful, as the matrix product $\mathbf{A}\: \mathbf{A}$ indicates whether technology $i$ receives an indirect spillover from technology $j$ via exactly one technology in between (in this case, the $i,j$ component of the matrix product is strictly positive) or not. Similarly, $\mathbf{A}^{k}$ indicates whether a technology is linked to another technology via exactly $k-1$ technologies in between. Given these points, we define
\begin{equation}\label{adjacency2}
\mathbf{A}^{\infty}:=\sum_{k=1}^{N}\mathbf{A}^{k}.
\end{equation}
This matrix $\mathbf{A}^{\infty}$ shows whether a technology is connected directly or indirectly to another technology. Based on this matrix, we introduce some spillover structures that will be of particular importance in our analysis.
\begin{definition}\label{defopt}\ \\
\begin{enumerate}
\item Technologies are \emph{independent}, if $\mathbf{F}$ and thus $\mathbf{A}^{\infty}$ are diagonal matrices.\footnote{If $\mathbf{F}$ is diagonal, each $\mathbf{F}^k$ is diagonal and thus the same holds for $\mathbf{A}^{\infty}$.}
 \item Spillovers are \emph{one-way}, if $\mathbf{F}$ (possibly after reordering technologies) is a lower triangular matrix.  In this case, $\mathbf{A}^{\infty}$ is also lower triangular.
 \item Technologies are \emph{separated} in two technology clusters, if $\mathbf{A}^{\infty}$ (possibly after reordering technologies)  can be written as 
 \begin{equation}\begin{pmatrix}\label{eq.separated}
A_1&0\\
0&A_2\\
\end{pmatrix},
\end{equation}
with $A_1$ and $A_2$ being square submatrices.
% \item Spillovers are \emph{symmetric}, if $\forall i,j\in N:\; F_{ij}=F_{ji}$.
 \item Technologies are \emph{strongly connected}, if all elements of $\mathbf{A}^{\infty}$ are strictly positive. A special case of strongly connected technologies are \emph{homogeneous} spillovers, where $\forall i,j \in N: F_{ij}=f>0$.
  \end{enumerate}
 \end{definition}
 The above characterisation captures some important cases. For independent technologies, we have at most intra-technology spillovers. The other extreme is the case of strongly connected technologies, where each technology influences each other technology either directly or indirectly. In between, we have the cases of one-way spillovers and separated technology clusters, which lead to some particularly interesting results, as our analysis will show. 
 
  Note that the above classification of spillover structures is loosely connected to the differentiation between reducible and irreducible matrices, which is central to the innovation network literature. If $F$ is irreducible, it cannot be one-way. However, if $F$ is reducible, this does not necessarily imply that the spillover structure is one-way, rather it indicates only that the structure  contains at least one cluster of technologies that is separated from all others.

These definitions can be used to show how different types of spillover structures lead to different growth prospects and patterns. The next proposition yields some first insights in this regard. 
\begin{proposition}\label{cor2}
Assume $\alpha>0$. Then:
\begin{enumerate}
	\item 	If there are no spillovers, that is $\mathbf{F}=0$, every technology grows linearly with long-run growth rates $\lim\limits_{t\rightarrow\infty}g_{q}:=\frac{\dot{q}}{q}$ tending to zero.
	\item If there are only intra-technology spillovers, that is, the technologies are independent, but we have at least one technology $i$ with $F_{ii}>0$,  the technology/technologies with the highest intra-technology spillovers will grow exponentially in the long run and all other technologies will stagnate.\footnote{Formally, there exists a non-empty set $\mathcal{S}\subseteq\{1,\ldots,N\}$ of all technologies $i$ with $F_{ii}=\max_{j\in \{1,\ldots,N\}}F_{jj}$. All technologies $i\in\mathcal{S}$ will grow at a constant rate in the long run, whereas all technologies $j\notin\mathcal{S}$ will stagnate.} 
	\item If there are no intra-technology spillovers ($F_{ii}= 0$ $\forall i=1,\ldots,N$), then:
\begin{enumerate}
\item If cross-technology interactions are one-way, there is linear growth in the long run. 
\item If there is at least one two-way cross-technology interaction (i.e., $\exists\: i\neq j:\; F_{ij}\: F_{ji}>0$), there is exponential growth in the long run. 
\end{enumerate}
\item  If all technologies are strongly connected, all technologies will grow exponentially in the long run. If the technologies are even homogeneous ($F_{ij}=f>0$, $\forall i,j$), the long-run growth rate will, in addition, be proportional to $f$.
\end{enumerate}
\end{proposition}
\begin{proof}
See Appendix \ref{proof2}
\end{proof}
This proposition shows how the structure of spillovers shapes growth prospects in our model. The first case establishes a benchmark: Without any spillovers, there can be growth but only at a declining rate. This is intuitive, as we have a constant number of scientists, which are the main resource for r\&d, and without spillovers the marginal productivity of a given number of scientists does not change over time, as existing knowledge does not render research more productive in this case.

The other extreme is the case of strongly connected spillovers, which can be seen as a second benchmark. In this case, each technology influences each other technology. Thus knowledge gained from one research activity is always increasing the productivity of all other research activities in later periods. This setting provides the strongest feedback effects that are possible in our model and thus, unsurprisingly, leads to the standard case of all qualities growing at identical and constant rates in the long run. Formally, the outcome of this case is easily observable from Eq. \eqref{eq.genrda}: By Lemma \ref{lem:1}, the long-run distribution of scientists (as described by the matrix $\mathbf{S}$ in Eq. \eqref{eq.genrda}) is constant and thus the r\&d growth engine is described by a set of differential equations that are all linear in the qualities $q(t)$.

The important point of Prop. \ref{prop2a} is, however, that there can be additional cases. Of particular interest is Case 3, which exemplifies the importance of the structure of spillovers: If we have two-way spillovers between technologies, we get exponential growth, albeit possibly only for some technologies. In contrast, one-way spillovers, where a technology receives spillovers only from 'upstream' technologies and causes spillovers only to  'downstream' technologies will not provide an outcome that deviates substantially from the case of no spillovers in the long run.

To interpret this, it is helpful to have a look at several special subcases of Case 3.
\begin{corollary}\label{cor2a}
Under the assumptions of Case 3 of Prop. \ref{cor2}, there is no exponential growth if each technology either causes or receives spillovers but never both, that is, for each technology $i$: if $F_{ij}>0$ for some $j\neq i$, then $F_{ji}=0$ for all $j\neq i$, and if $F_{ji}>0$ for some $j\neq i$, then $F_{ij}=0$ for all $j\neq i$.
\end{corollary}
\begin{proof}
Under the assumptions of the corollary, the technologies can be re-ordered, so that the matrix $\mathbf{F}$ has a lower triangular structure (see Def. \ref{defopt} Case 2).
\end{proof}

\begin{corollary}\label{cor2b}
Under the assumptions of Case 3 of Prop. \ref{cor2}, exponential growth is feasible if and only if there exists a 'core' of technologies, i. e. at least a pair of technologies positively affecting each other (circular chain).
\end{corollary}
\begin{proof}
If there are no circular chains of interaction, the matrix $\mathbf{F}$ can always be brought to a triangular structure by re-ordering the technologies. Thus if this is not the case, exponential growth is possible.
\end{proof}
Proposition \ref{cor2} and the above corollaries show that the structure of cross-technology interactions strongly influences the   prospects of long-run growth. Specific structures of cross-technology interactions are required to induce exponential growth, irrespective of the strength of the interactions.

Case 3 (a) of	Prop. \ref{cor2} shows for which structures of interactions there is no chance for exponential growth.  As Corollary \ref{cor2a} suggests, this case contains all interaction structures, where each technology either only contributes to the development of other technologies or only profits from the state of other technologies but never both simultaneously. In such cases, r\&d can induce development but only with decreasing growth rates. Thus a general purpose technology (GPT), as in \cite{Bresnahan1995}, is not sufficient to warrant exponential growth: The GPT itself has to receive a positive feedback, either from its own development or from the other technologies.

Corollary \ref{cor2b} highlights that to achieve constant positive growth rates, it is necessary to have at least one circular chain of development. This is similar to the concept of a core in \cite{rebelo1991}. Thus exponential r\&d-driven growth requires two-way interactions, which, however, do not necessarily have to be between two technologies. A chain of development, where each technology in the chain has a positive impact on the next technology and where there is a positive spillover from the end of the chain to its beginning (so that the chain is 'circular') also suffices, as does an intra-technology spillover (which is the shortest possible chain of development), as described by Case 2 of Prop. \ref{prop2a}.

These results are fairly intuitive. Quality grows at most linearly over time in the absence of spillovers (see Case 1 of Prop. \ref{cor2}). If spillovers are one-way, some technologies might receive a positive spillover from other technologies. But as these spillover-generating technologies have themselves only linearly growing quality, the spillover cannot induce exponential growth (the effect of the spillover increases only linearly with time). In such a case, the dynamic system Eq. \eqref{eq.genrda} can be ordered in a way so that the equation for each technology $i$ depends only on qualities $j<i$, that is, there is no equation where the growth of a technology $i$ is influenced by the quality of this technology itself or by the quality of some technology that is influenced by this technology $i$.

Only the presence of feedback effects, where some technology $i$ positively influences another technology $j$ and, in turn, gets a positive  spillover from technology $j$, can generate exponential growth. In such a setting, technology $i$ has not only an effect on the other technology $j$, but (via the bidirectional spillover) its own growth causes an indirect spillover to itself. This is possible either by a bidirectional interaction between two technologies or by more complex circular chains of innovation, as described in Corollary 2. This can be seen as a generalization of the 'common pool of knowledge'  type of spillover that is used in many growth models; such a knowledge pool is not required to generate exponential growth, but the  aspect that at least one technology has an indirect positive effect on itself via a circular chain of spillovers is necessary.

Note that Prop. \ref{cor2}  relies solely on the structure of spillovers, not their size.  If, for example, the structure conforms to the case of one-way spillovers, even arbitrarily strong spillovers between technologies will not suffice to induce exponential growth.

Finally, Case 2 of Prop. \ref{cor2}  shows the relative importance of knowledge or spillover-driven growth and individual technologies' growth.  Positive intra-technologies spillovers can easily generate exponential growth and this is well-known in the growth literature. 

\begin{figure}
	\subfloat[Technology levels]{\includegraphics[scale=0.35]{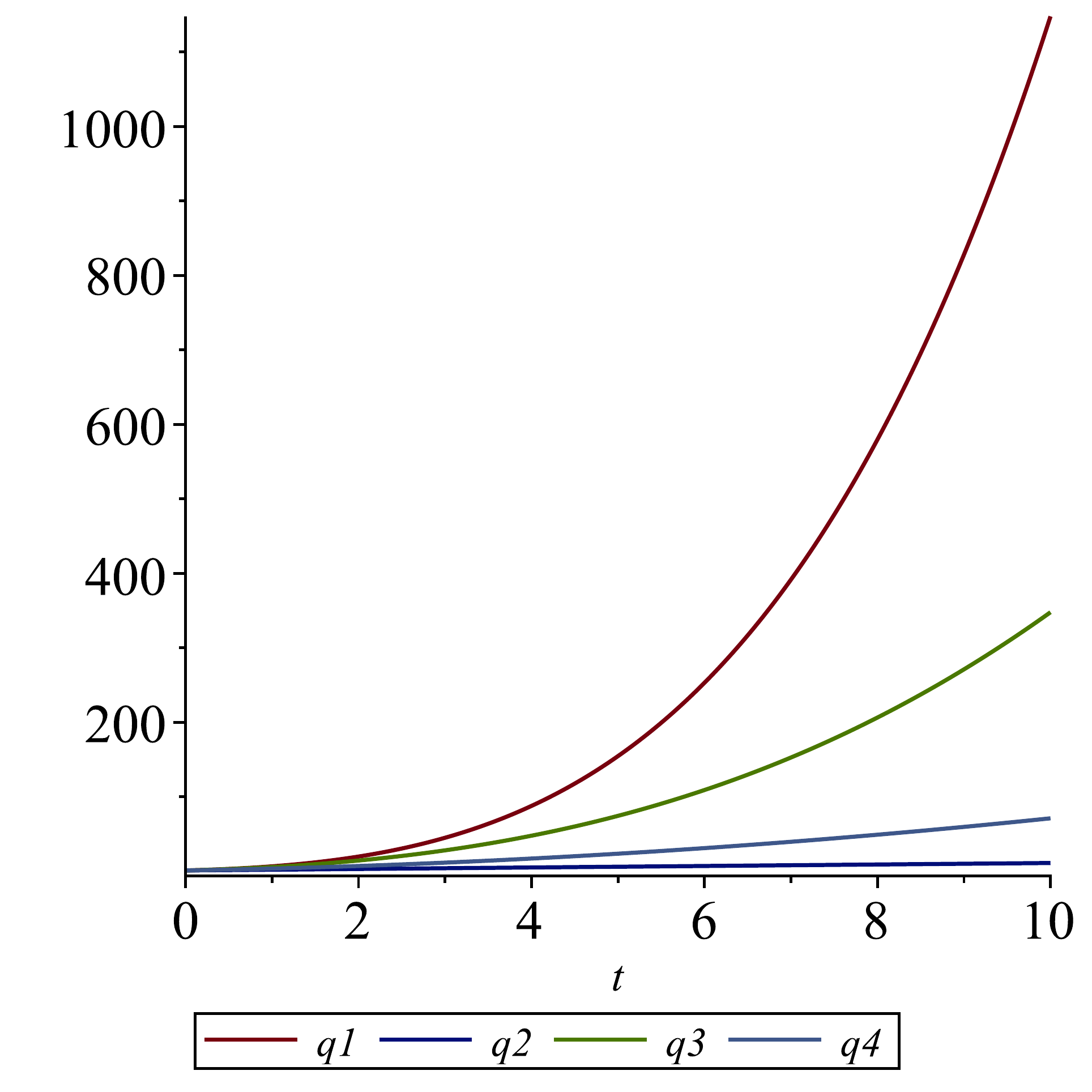}}
	\subfloat[Growth rates]{\includegraphics[scale=0.35]{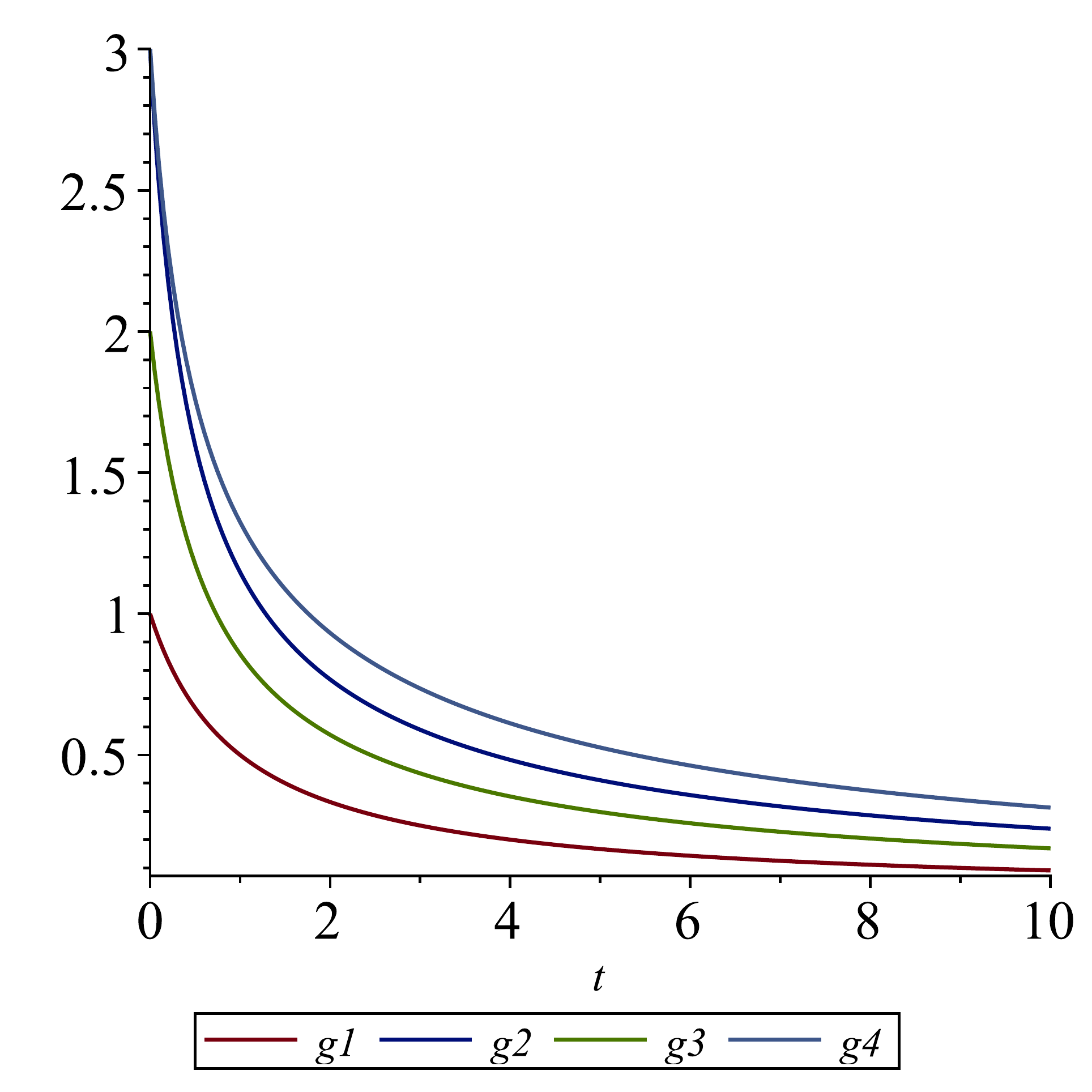}}
	\caption{Less than exponential growth}\label{polgro}
\end{figure}

An illustrative example can be given for $N=4$ by choosing
  \begin{equation}
\mathbf{F}=\begin{pmatrix}\label{eq.onewayex}
0&0&0&0\\
1&0&1&0\\
1&0&0&0\\
1&1&0&0
\end{pmatrix}.
\end{equation}
 This matrix resembles a setting where technology $1$ only gets spillovers, technology $4$ only causes spillovers, and technologies $2$ and $3$ interact with the  other technologies in a way  so that no circular chain of spillovers exists.\footnote{Indeed, this setting can be reordered to a one-way spillover structure with   \begin{equation*}
\mathbf{F}=\begin{pmatrix}
0&1&0&1\\
0&0&1&1\\
0&0&0&1\\
0&0&0&0
\end{pmatrix},
\end{equation*}
by switching the numbering of Technology $1$ and $4$.} Thus, such a spillover structure cannot induce exponential growth. Formally, this matrix has all eigenvalues strictly equal to zero and thus only less-than-exponential growth is possible. Technology levels and growth rates are illustrated in Figure \ref{polgro}.
 This example illustrates a situation where, initially, some technologies can grow faster than linearly. But the competition for scientists eventually drives out the development of all technologies except the fastest growing one. This, in turn, reduces the growth of the fastest growing technology, as it does not receive increasing spillovers from the other technologies anymore. Consequently, growth is reduced to linear growth in the long run. 
 
\begin{figure}
	\subfloat[Technology levels]{\includegraphics[scale=0.35]{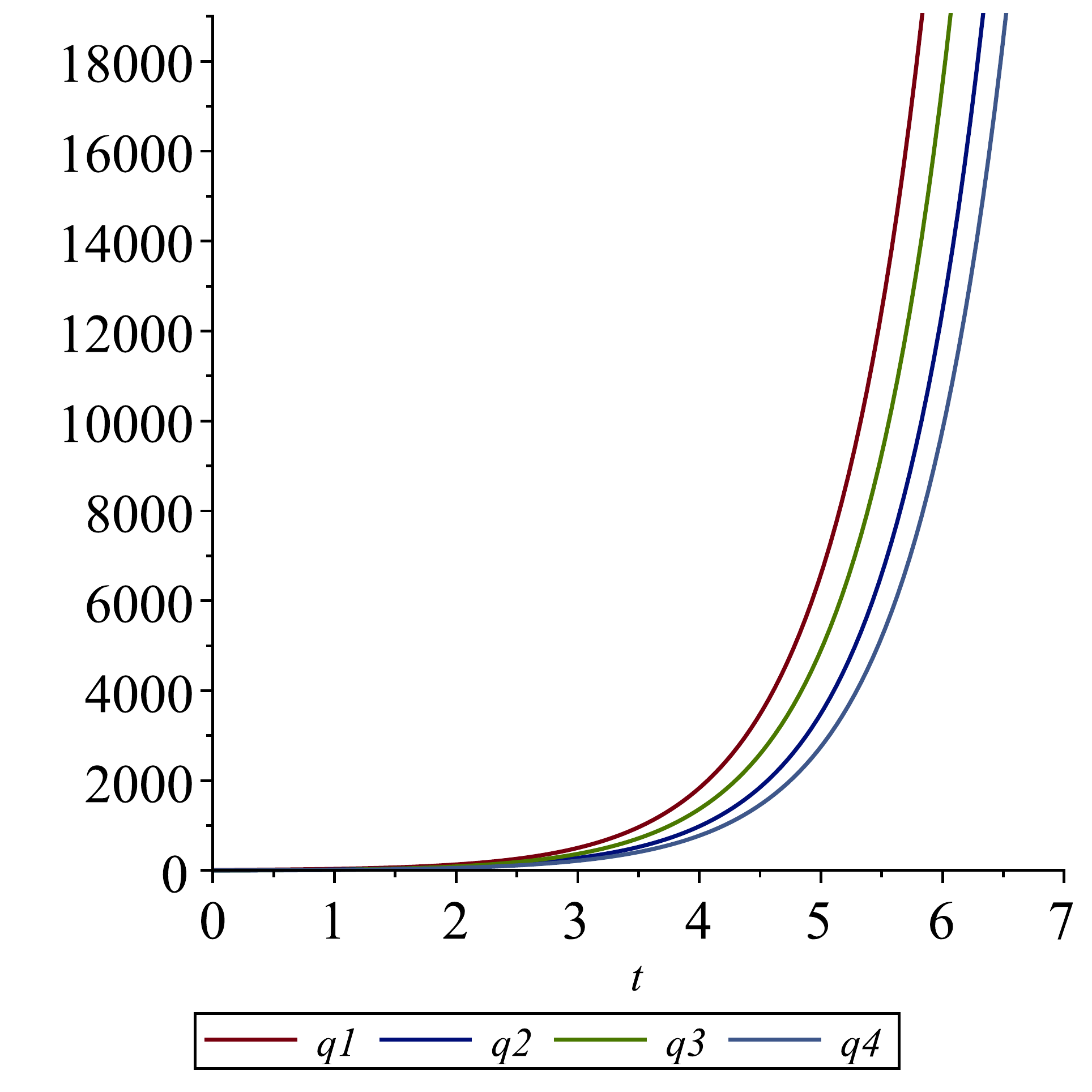}}
	\subfloat[Growth rates]{\includegraphics[scale=0.35]{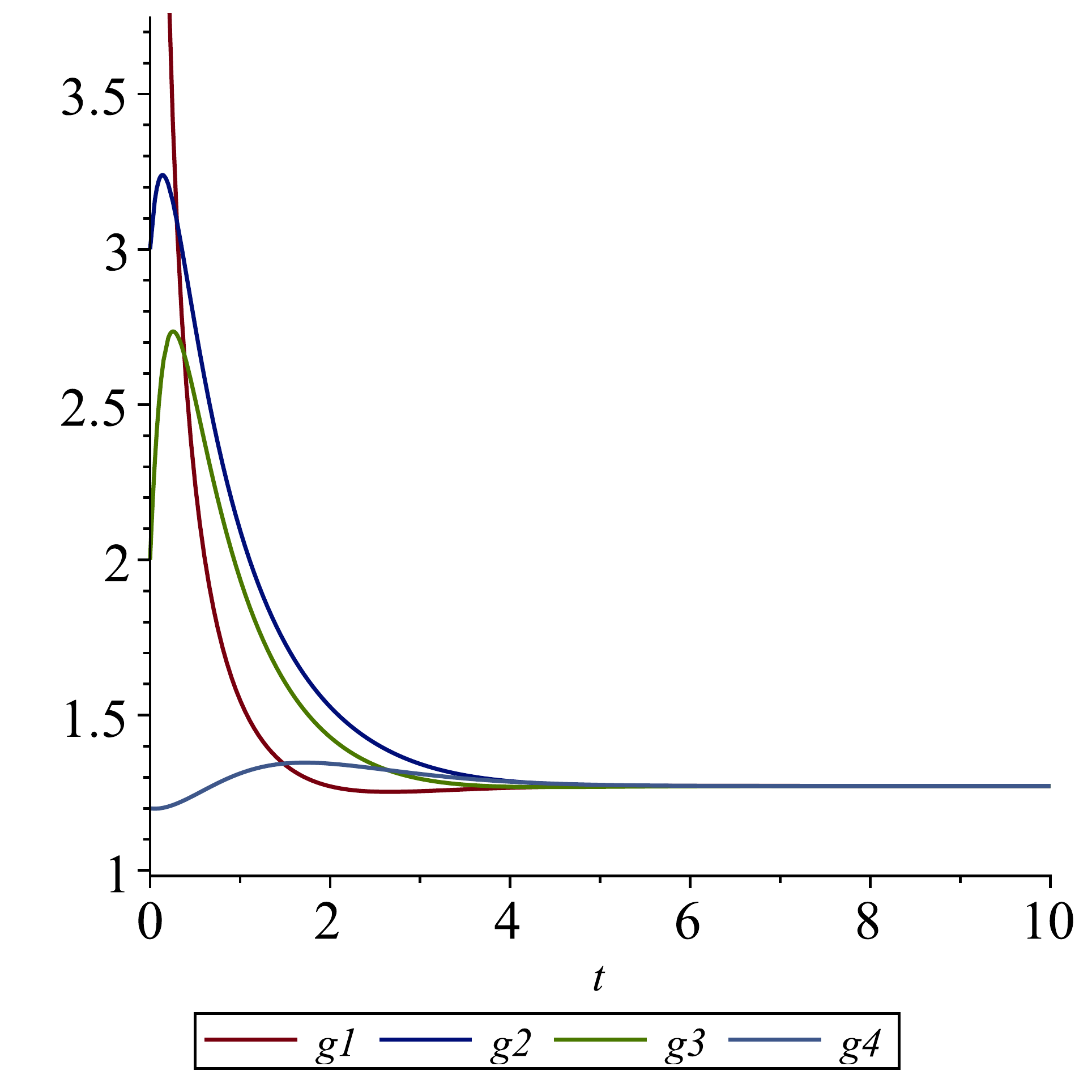}}
	\caption{Exponential growth pattern}\label{expgro}
\end{figure}

 However, if we modify the example so that 
   \begin{equation}
\mathbf{F}=\begin{pmatrix}\label{eq.onewayex2}
0&0&0&1\\
1&0&1&0\\
1&0&0&0\\
1&1&0&0
\end{pmatrix},
\end{equation}
we get a circular chain of interactions from Technology $1$ to Technology $4$ to Technology $2$ and then back to Technology $1$. This creates the opportunity for exponential growth, as an improvement of Technology $1$ generates an indirect spillover (via Technologies $4$ and $2$) to itself. Again, formally this matrix has at least one positive eigenvalue, which leads to an exponential solution for \eqref{eq.genrda2}. This type of growth is illustrated in Figure \ref{expgro}.

Relating our results so far to the literature on innovation networks shows that an irreducible spillover matrix can induce exponential growth, as such a matrix describes a strongly connected directed network of spillovers. Moreover, this growth will spread out to all technologies, as an irreducible spillover matrix admits no separated technologies or separated technology clusters. 

On the contrary, if $F$ is reducible but contains at least one circular chain, there can be exponential growth due to spillovers but it will be concentrated in some specific cluster (a 'core') while other sectors/technologies will not grow beyond a steady state ('periphery'), as they will not be able to attract scientists in the long run. Such a situation is observed in some modern economies with regard to the spatial distribution of growth (see \cite{storper2011} and the related economic geography literature). Furthermore, \cite{Acemoglu2016} suggest that spillovers are concentrated on some key industries, which could easily lead to a similar effect.

To link our results to the network literature we re-state a part of our main result in terms of the irreducibility of $F$:
\begin{corollary}\label{prop2a} Assume Lemma \ref{lem:1} holds. Then:
	\begin{enumerate}
		\item If  $\mathbf{F}$ is irreducible (for example, all technologies are strongly connected) all technologies survive in the long run and grow exponentially
		\item If $\mathbf{F}$ is reducible (for example, there are only one-way spillovers), not all technologies may survive in the long run and growth may be polynomial.
	\end{enumerate}
\end{corollary}
\begin{proof}
	By assumption, the matrix $\mathbf{F}$ is non-negative. Therefore, the proof follows from the Perron-Frobenius theorem\footnote{See, for example, \cite{BENVENUTI2014}}: If a non-negative matrix is irreducible, its maximum eigenvalue is strictly positive and real, implying an exponential growth of all technologies. If $\mathbf{F}$ is reducible, its maximal eigenvalue may be zero (but real), implying a linear trend in the solution of \eqref{eq.genrda2}. As we do not exclude cross-technology spillovers, this linear trend can be carried over to other technologies as a polynomial one. 
\end{proof}
In addition, it is possible to add some insights to Prop. \ref{prop.lrgrowth}, for the case of a reducible spillover matrix.
\begin{corollary}
	Let $\mathbf{F}$ be block-diagonal (and thus reducible) with $1<S< N$ blocks. Additionally, let there be no intra-technology spillovers ($\forall i\in N:\;F_{ii}=0$). Then:
	\begin{enumerate}
		\item If all technologies start with the same initial quality ($\forall i:\; q_{i}(0)=q$), only the block with maximal overall spillovers continues to grow in the long run.
		\item If $\exists! i_{0}:\;q_{i_{0}}(0)>q_{i}(0)$ and all non-zero spillovers are the same in size, only the block containing technology $i_{0}$ continues to grow in the long-run.
	\end{enumerate}
\end{corollary}
\begin{proof}
	\begin{enumerate}
		\item If all technologies start equally, each block initially attracts the same labour. However, as spillovers in one block benefit the technologies more, they eventually get an advantage over other blocks and will attract all scientific labour (see Lemma \ref{lem:1}).
		\item Once there are no advantages in terms of spillovers, all blocks start growing at the same speed, but the headstart technology $i_{0}$ will benefit its block more than others, thus eventually attracting all labour to this block (again see Lemma \ref{lem:1})
	\end{enumerate}
\end{proof}
This result shows how spillovers and initial conditions interact for the case where there are several separated clusters of technologies that compete for scientists. If all clusters have the same spillovers, then the cluster with an initial advantage will be the one that continues to grow. If all clusters have the same initial starting point in terms of qualities, then the cluster with the highest spillover will win the competition for scientists.

As a final step of our analysis, we now turn to the question how complex growth can be. First, we ask  whether growth will be monotonic or whether there might be fluctuations. The following corollary provides an answer for the long run. 
\begin{corollary}
	If technology-driven long-run growth is present, it is always  monotonic in the long run.
\end{corollary}
\begin{proof}
	In the long-run, the allocation of scientists is constant across technologies and, by the Perron-Frobenius theorem, the maximal eigenvalue of a non-negative matrix is non-negative and real. This implies monotonic growth for $t\rightarrow\infty$.
\end{proof}
This result is not very surprising: We consider a case with only positive spillovers; a technology's development cannot possibly become less easy due to improvements in other technologies. Thus, fluctuations could only stem from a reallocation of scientists. But, in the long run, the allocation of scientists does not change anymore, implying that the long-run development cannot exhibit fluctuations.

This result shows that growth patterns will, eventually, become rather simple. In the long run, growth will be either exponential or linear with all technologies that do not stagnate growing at identical rates. Given that this holds for all possible spillover structures (apart from the case of negative spillovers that we did not cover so far), this is a rather strong result. However, it only applies to the long run. As the following result shows, more complex dynamics are feasible for particular spillover structures in the short run.

To provide an example of such conditions, consider a setting with two disjoint clusters of technologies $\mathcal{I}_1,\mathcal{I}_2\subseteq \{1,\ldots,N\}$. Assume that the technologies within each cluster are strongly connected, in the sense of Definition 1, but that there is only a one-way connection between these clusters, that is, at least one technology from $\mathcal{I}_2$ receives a spillover from at least one technology in $\mathcal{I}_1$ but there is no spillover from a technology in $\mathcal{I}_2$ to a technology in $\mathcal{I}_1$. In total, this amounts to a spillover structure that can be characterized by 
\begin{equation}\mathbf{A}^{\infty}=\begin{pmatrix}\label{eq.one-way}
A_1&0\\
A_2&A_3\\
\end{pmatrix},
\end{equation}
where all submatrices $A_1,A_2,A_3$ are square matrices with all elements of these matrices being strictly positive.\footnote{Note that even if there is only a single spillover between technologies in $\mathcal{I}_1$ and $\mathcal{I}_2$, the fact that these subsets are strongly connected implies that each technology in $\mathcal{I}_2$ is at least indirectly connected to each technology in $\mathcal{I}_1$. Thus, $A_2$ cannot have elements that are zero.}

In such a situation, there can be a technology transition. That is, at the beginning the technologies in $\mathcal{I}_1$ will be developed, but eventually development will shift completely to the technologies in $\mathcal{I}_2$. The following lemma states this formally.

\begin{proposition}\label{lem:sr1}
Assume that the spillover structure is characterized by Eq. \eqref{eq.one-way} with $A_1$ being an $N_1\times N_1$-matrix and all elements of $A_1,A_2,A_3$ being strictly positive. Then, there exist $\phi_1,\phi_2>0$, so that multiplying all $F_{ij}$ by $\phi_2$ for all $i,j>N_1$ and by adding $\phi_1$ to all $q_k(0)$ for $k\in\{1,\ldots,N_1\}$ leads to a growth pattern where the technologies $1\ldots,N_1$ attract more scientists in total than the remaining technologies for $t=0$ (and are thus the initially most strongly developed technologies), but that, for $t\rightarrow\infty$, only the technologies $N_1+1\ldots,N$ are developed.
\end{proposition}
\begin{proof}
As we have two-way connections between technologies, Prop. \ref{cor2} implies that there is exponential growth in the long run. Thus, Lemma \ref{lem:1} can be applied, showing that all technologies in $\mathcal{I}_2$ will always grow in the long run, as they receive a spillover from the technologies in $\mathcal{I}_1$ and as the technologies within both subsets are strongly connected. By multiplying all $F_{ij}$ by a sufficiently large $\phi_2>0$ for $i,j>N_1$, we can ensure that only the technologies in the subset $\mathcal{I}_2$ attract scientists (and thus grow) in the long run, as the technologies in $\mathcal{I}_1$ receive no spillover from the technologies in $\mathcal{I}_2$. Note that for a sufficiently large $\phi_2$ this holds for a wide range of initial conditions for technologies in $\mathcal{I}_1$ (again, due to the one-way spillover structure between these technology clusters).

In contrast, Eq. \eqref{shares.1} implies that for $t=0$, we can increase the share of scientists working on technologies in the set $\mathcal{I}_1$ by increasing the initial qualities of these technologies compared to those of the technologies in  $\mathcal{I}_2$. Thus, we can find a $\phi_1>0$, so that adding $\phi_1$ to all $q_k(0)$ for $k\in\{1,\ldots,N_1\}$ leads to more scientists being attracted to technologies in the set $\mathcal{I}_1$ for $t=0$, which implies that these technologies will initially grow more strongly than the technologies in the set $\mathcal{I}_2$.
\end{proof}
This final result indicates that spillovers can induce rather complex growth patterns in the short run. The reason is, again, the interaction between spillovers and the competition for scientists. The setting referred to in Prop. \ref{lem:sr1} resembles a situation where we have two clusters of technologies with strong connections within each cluster and where one cluster supports growth of the other, but not vice versa. In such a situation, it is clear from the result so far that the spillover-receiving cluster will always grow in the long run. What Prop.  \ref{lem:sr1} adds is that, based on such a structure, it is always possible to generate a case where only this cluster will grow in the long run but, initially, growth is driven by the other cluster.

This is of particular interest for describing the development in some sectors, where transitions between technologies have been observed in the past. A prime example is the energy sector, where we have seen a transition from biomass (the main energy source before the industrial revolution) to coal to oil and, recently beginning, to renewable energies (and, possibly, nuclear energy), see, for example, \citet{Sovacool}.

\begin{figure}
	\subfloat[Share of scientists for the different technologies]{\includegraphics[scale=0.75]{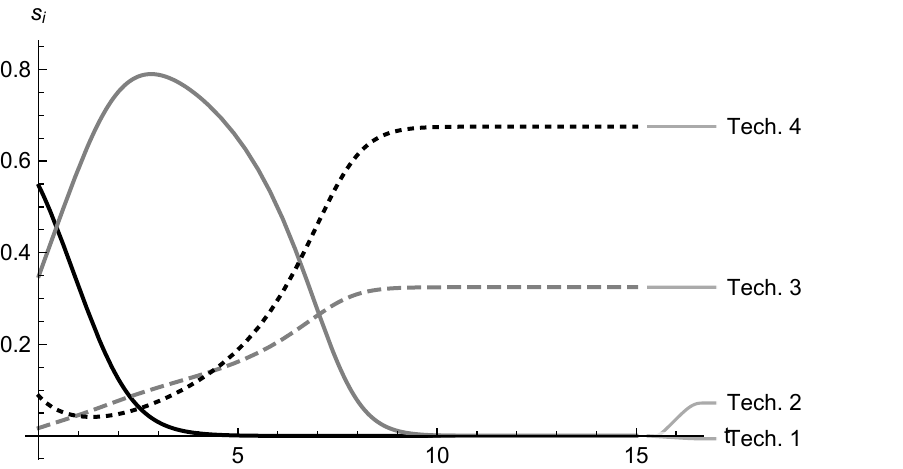}}
	\subfloat[Growth rates of individual technologies]{\includegraphics[scale=0.75]{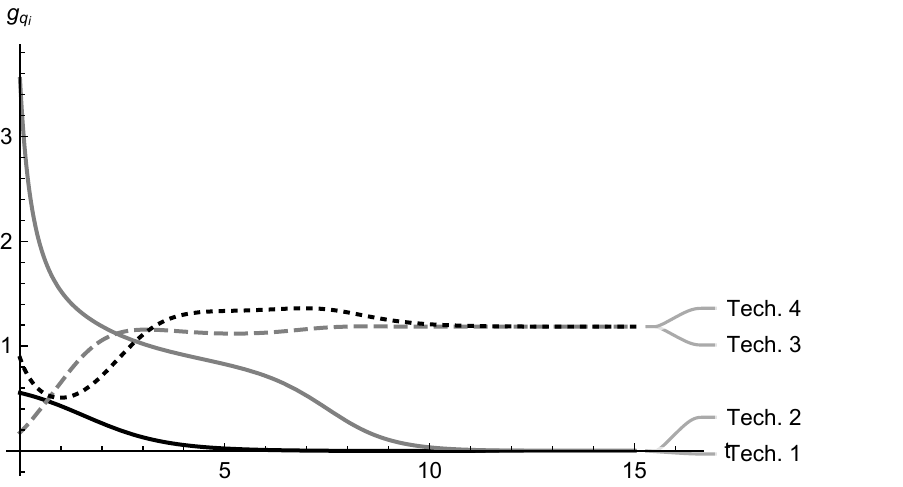}}
\hspace{0mm}
\centering		\subfloat[Growth rate in the labour-intensive sector]{\includegraphics[scale=0.5]{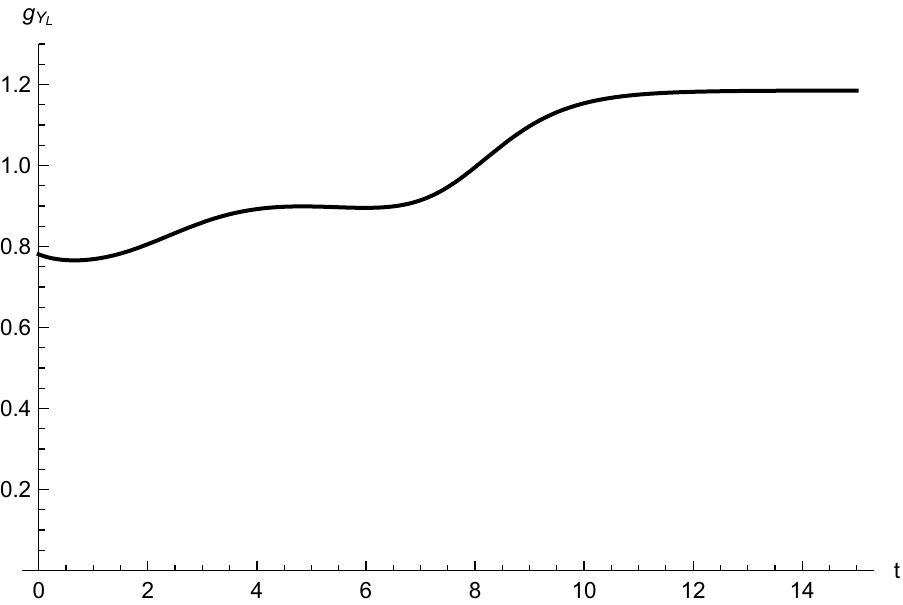}}
	\caption{Two subsequent technology transitions, calculated for the spillover structure shown in Eq. \eqref{eq.complexgrowth} with initial values $q_1(0)=1$, $q_3(0)=q_3(0)=q_4(0)=1/10$ for $\nu=1/2$, $\alpha=0$, and $S=1$.}\label{complexgrowth}
\end{figure}

Note that the technology transitions described in the above proposition cannot only occur between two but between several blocks of technologies. Consider, for example, the following spillover structure with $N=4$ technologies, where we have growth that shifts from Technology $1$ to Technology $2$ and then to joined development of Technologies $3$ and $4$:
   \begin{equation}
\mathbf{F}=\begin{pmatrix}\label{eq.complexgrowth}
3/4&0&0&1\\
1/2&1/2&0&0\\
0&1/3&0&1\\
0&0&3&0
\end{pmatrix}.
\end{equation}
Figure \ref{complexgrowth} shows the resulting growth pattern. First, Technology $1$ drives growth, but as it has only a small intra-technology spillover and receives no spillovers, r\&d for this technology is only high due to its good initial starting position and declines rapidly once some of the other technologies become more developed. In a second phase, Technology 2 takes over, as it is stimulated by a spillover from Technology 1 and has a higher intra-technology spillover. But eventually, a spillover from Technology 2 to Technology 3 leads to the development of the latter technology, which forms a spillover chain with Technology 4 that provides a strong growth engine. 

As can be seen from this example, the shift of scientists among different technologies leads to a rather complex growth pattern: The allocation of scientists changes non-monotonically, as do the growth rates of the individual qualities. Furthermore, this leads to a growth rate in the labour-intensive sector that switches, with each transition, between different levels.\footnote{Note that these switches will always lead to higher growth rates, as scientists shift only to technology clusters with better growth prospects.}

\section{Extensions}\label{sec:ext}
As we analyse a comparatively complex setting ($N$ intertemporally optimizing r\&d firms connected via heterogenous spillovers), we have to use some restrictive assumptions to gain insights. However, some of these assumptions can be partly relaxed, others have arguably only limited importance for our results. In this section, we briefly provide some extensions to our results and discuss the importance of the strongest assumptions used in our analysis.

One assumption that can be partly relaxed is the assumption of non-negative spillovers. This assumption is mainly required to avoid corner solutions, which are hard to handle in high-dimensional dynamic models. 

\begin{figure}
	\subfloat[Technology levels]{\includegraphics[scale=0.35]{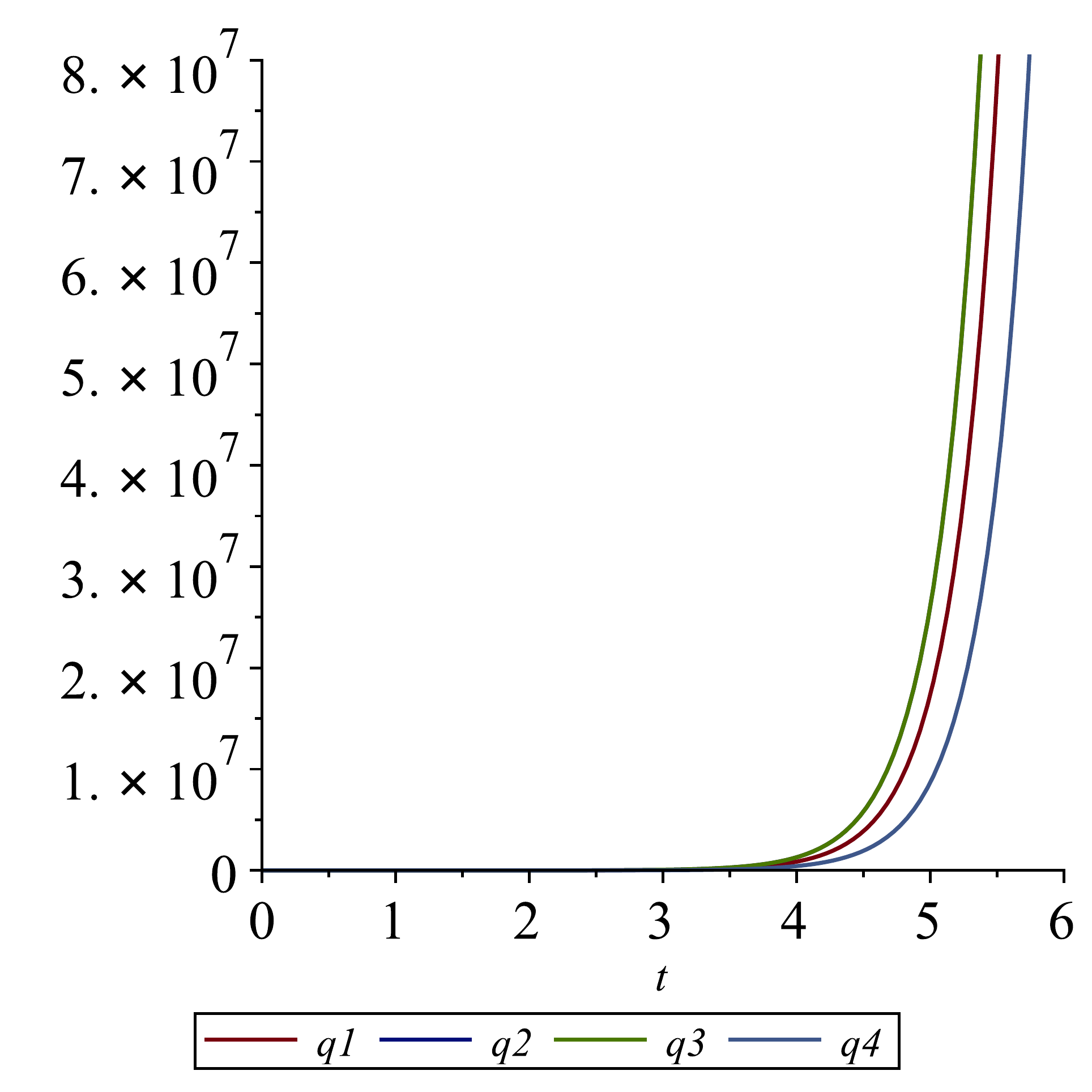}}
	\subfloat[Growth rates]{\includegraphics[scale=0.35]{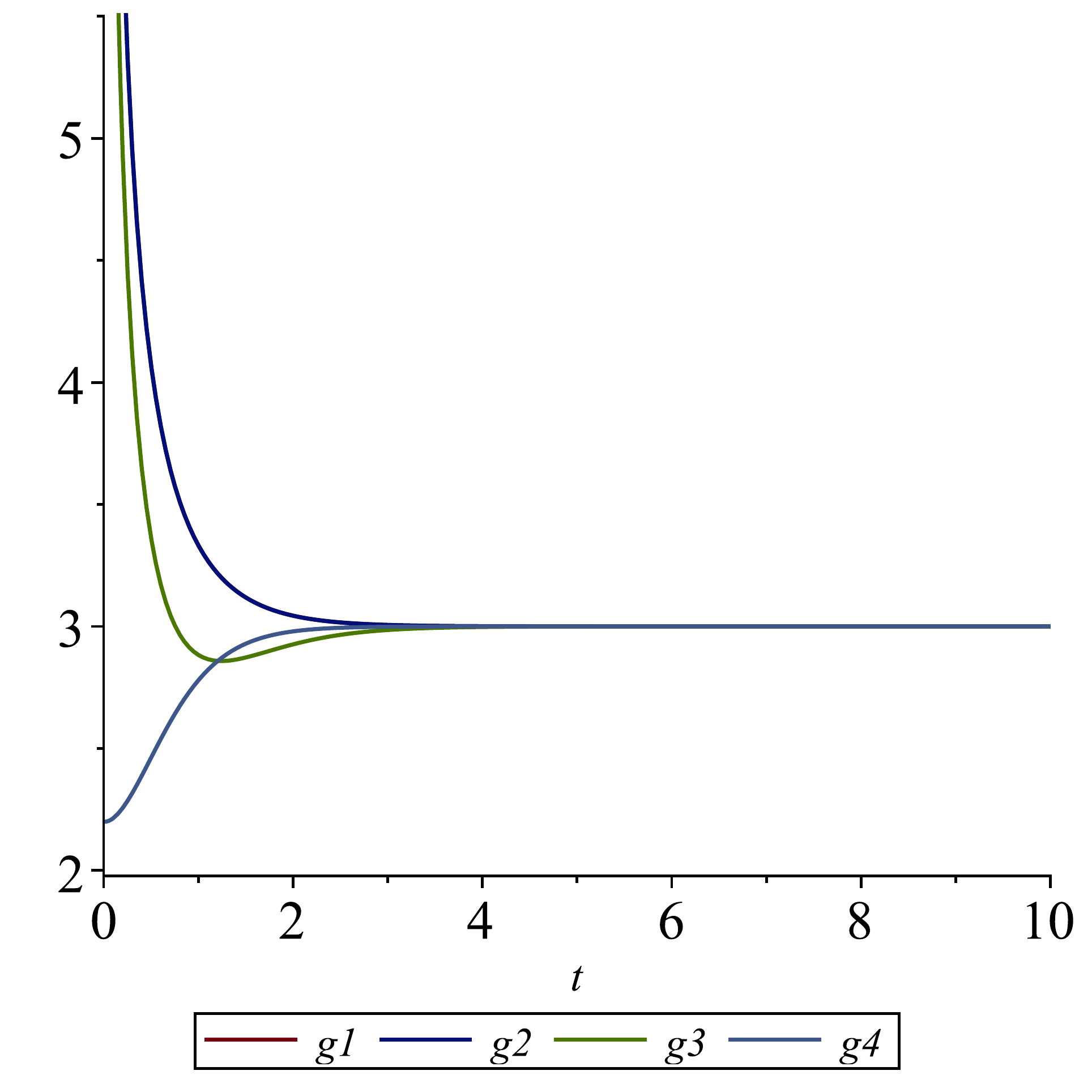}}
	\caption{Technology growth with eventually nonnegative $F$}\label{noexp}
\end{figure}

It is possible to generalize our results to some cases of negative cross-technology externalities. First, assume that $\mathbf{F}$ contains some negative spillovers but that Lemma \ref{lem:1} still holds, that is, the technologies in the set $\mathcal{I}_{\infty}$ enjoy nonnegative overall spillovers. Such a case occurs in two settings: First, if negative spillovers stem only from technologies whose development will eventually cease (those in the set $\mathcal{I}_{0}$), and, second, if negative spillovers stem from technologies that continue to grow but these spillovers are always dominated by positive spillovers from technologies in $\mathcal{I}_{\infty}$.\footnote{As all technologies grow proportionally to each other in the long run, this dominance can be easily verified.} To analyse this case, we need the following definition:
\begin{definition}
	The matrix $\mathbf{F}$ is eventually non-negative, if there exists $k\in\mathbb{N}$ so that $\mathbf{F}^{k}\geq 0$.
\end{definition}
Thus, an eventually non-negative matrix may contain negative entries, but some power of such a matrix is a non-negative matrix. Using this definition, we can state the following extension to our results.
\begin{corollary}\label{cor.ext1}
	Let $\mathcal{I}_{\infty}\neq \emptyset$. If $\mathbf{F}$ contains negative entries but is eventually non-negative and $\forall i\in\mathcal{I}_{\infty}:\; s^{\infty}_{i}>0$, Propositions \ref{prop.lrgrowth} and \ref{prop2a} as well as Corollaries 1-3 hold.
\end{corollary}
\begin{proof}
	This follows from the definition of an eventually non-negative matrix: The long-run dynamics of the system depend on the matrix $\mathbf{F}$ only (as long as there exists a long-run set of growing technologies). The theory of dynamic systems shows that the solution to \eqref{eq.genrda2} has the form $\mathbf{q}=\mathrm{e}^{\mathbf{F^{\infty}}t}+C$ and that the matrix exponent is non-negative as long as $\mathbf{F}$ is eventually nonnegative, see, e. g. \cite{Naqvi2003}. Thus, we are back in the case analysed in the preceding section with regard to all results that apply only to the long run.\end{proof}
As an example, consider
\begin{align}
\mathbf{F}=\begin{pmatrix}
1&1&1&1\\
1&1&1&1\\
-1&1&1&1\\
1&0&1&1
\end{pmatrix}.
\end{align}
This spillover matrix contains a negative impact of Technology $1$ on Technology 3 and otherwise non-negative entries. Direct computation verifies that $\mathbf{F}^{2}$ is non-negative, so that $\mathbf{F}$ is eventually non-negative. Figure \ref{noexp} illustrates the quality development and growth rates for this example, showing that the main results of the preceding section (as mentioned in Cor. \ref{cor.ext1}) hold despite the negative spillover.
y

A second assumption of our analysis is that  we have assumed that the structure of spillovers remains unchanged over time. Again, our results can be slightly generalized. Assume that $\mathbf{F}$ is dynamic, that is, $\mathbf{F}=\mathbf{F}(t)$, but that the number of technologies does not change over time. Then, if the main structure of spillovers does not change (i.e., all zeros and signs of entries in $\mathbf{F}(t)$ are invariant w.r.t. $t$) and if $\mathbf{F}(t)$ converges to some constant matrix in the long run, it is obvious that all results in the preceding section that depend on the structure, not on the relative sizes of the spillovers, will still apply.

Third, with some minor adjustments, our framework can be extended to a continuous  range (rather than countable number) of technologies. To this end all 'matrices' have to be replaced by operators and the main results hold for particular classes of such operators.  For example, Prop. \ref{prop.lrgrowth} remains valid  as long as the operator $\mathcal{F}$ is compact and bounded. Similarly, Prop. \ref{cor2} continues to hold, if the operator $\mathcal{F^{\infty}}$ is compact and of spectral class\footnote{See e. g. \cite{dunford1954} for a definition.} $m<\infty$, that is, it has only distinct eigenvalues and its complexity\footnote{This complexity is defined as the difference of the number of complex and real eigenvalues.}  is not infinite.

Finally, our analysis is constrained by the assumption $\beta=1/2$, which implies that the profits of the r\&d firms are linear in the quality of their technology. This assumption is essential to gain formal results. Without this assumption, the co-state variables would depend on the evolution of the qualities and the model cannot be solved anymore, except for highly specific spillover structures. 

However, investigating the changes to incentives implied by moving away from this linearity indicates that relaxing this assumption towards larger values of $\beta$  is unlikely to cause substantial changes to most of our results. To see this, note first that, in our model, the dynamics depend only on spillovers and on the allocation of scientists. As the spillovers result directly from the qualities, changes to $\beta$ can affect our results solely via a changed allocation of scientists.

For $1/2<\beta<1$, the profits in the r\&d sector will be a strictly concave function of qualities. This would imply a stronger incentive to allocate scientists more evenly to the development of the different technologies. However, compared to the linear spillover structure, this effect is small in the long run. To see this, compare the marginal profits (the derivates of $\Pi_{k}(q_k(t),t)$ as defined in Section \ref{sec.mod})  of two technologies $i$ and $j$ with qualities $q_i$ and $q_j$. Simple computations show that the quotient of these marginal profits is proportional to $(q_i/q_j)^{\frac{1-2\:\beta}{\beta}}$. For the case used in our analysis ($\beta=1/2$), this relation is simply a constant. For $q_i/q_j>1$ and $1/2<\beta<1$, which is the interesting case,  this relation changes less than proportionally with the relative qualities;  any effect that is linear in qualities, like the spillovers, will dominate. Thus, for $1/2<\beta<1$, there is an incentive to develop a technology that is currently less developed compared to other technologies, as the marginal profit is higher for this technology. But the differences in the marginal costs of development (due to differences in spillovers) will eventually dominate this effect.

Thus, even with strictly concave profits, some technologies can still stagnate. Furthermore, for the technologies that continue to grow, the above relation of marginal profits simply converges to a constant in the long run. Accordingly, we can expect that all results referring to the long run development will remain mostly unaffected. This holds in particular for Lemma \ref{lem:1}, Prop. \ref{prop2a}, as well as Corollaries 1-3. Proposition \ref{prop.lrgrowth} would hold with  modified vectors $F_i$ that compensate for the differences in marginal profits. As argued above, in the long run (to which the proposition refers), this compensation would only require constant factors.

Overall, these arguments indicate that most of our results can be transferred to the case $1/2<\beta<1$ and thus to a setting that is quite widely used in growth theory.

\section{Conclusions}\label{sec.concl}
Long-run growth is often considered to be the result of spillovers in r\&d: The development of one technology facilitates the development of others, generating increasing returns to scale and thus the possibility for continued growth. Empirical research has shown that such spillovers are not always uniform, that is, not every technology causes similar spillovers and not every technology benefits in a similar way from spillovers. However, most theoretical models are based on such a homogeneous spillover structure. In this paper, we have  studied how more general spillover structures between individual technologies influence the prospects for and characteristics of growth. To this end, we have advanced a detailed model of the r\&d sector and embedded this in a simple endogenous growth model.

Our results indicate that the structure of r\&d interactions is indeed important. Whereas some types of cross-technology interactions, like one-way spillovers, cannot possibly induce the exponential growth that is a typical feature of  growth models, other spillover structures can, such as intra-technology spillovers or circular chains of development. Furthermore, depending on the structure of spillovers, only some technologies might be developed in the long run and which technologies are developed can depend on initial conditions, implying some form of path-dependency. Furthermore, even though our model conforms to standard models in that, in the long run, the qualities of all technologies that are still developed will grow at the same rate, we have shown that some types of spillover structures can generate rather complex and interesting growth patterns in the short run. An example is what we have described as a technology transition, where the main engine of growth shifts between different clusters of technologies over time, which can lead to changing growth rates.

Our results complement studies on innovation networks, such as \cite{Acemoglu2012}, in that we do not investigate the distribution of growth or the propagation of shocks in an economy but rather study the question which spillover structures can lead to  what kind of aggregate growth patterns. This question has some relevance, as recent empirical studies have observed non-exponential growth patterns \citep{kogan2017,gordon2016,jones2017}.

Overall, our results have several implications. First, they  suggest that the assumption of homogenous spillovers that is often used in growth models generates dynamics that are among the most simple ones possible in that all technologies will see continued development and growth is always exponential. Given the complexity of innovation processes in many sectors, a perfectly homogeneous spillover structure appears to be  unlikely. Thus, it might be productive to consider other types of spillover structures in growth models. 

Second, using a wider range of spillover structures could facilitate the description and analysis of more complex growth patterns, such as transitions between different technology clusters.  Our results indicate what kind of structures could be helpful for explaining observed developments. Also, our theoretical results could be applied and tested empirically. Spillovers induce a correlation of quality developments and thereby a correlation of sectoral growth rates, which can be observed. A strong heterogeneity of the correlation coefficients between sectoral growth rates would thus indicate that the structure of spillovers could be complex and should be modeled in detail, whereas a high homogeneity of the correlation of sectoral growth rates would indicate that the widely used model of homogeneous spillovers could be a good approximation. 

Finally, our insights lead to an interesting question: If the structure of spillovers has a substantial influence on growth prospects and patterns, what  kind of technology base is most conducive for economic development? Should a country focus on sectors using similar technologies, which would most likely have two-way spillovers, or is it better to have circular chains of technological development, where developments of one sector are used in another ('downstream' ) sector with few 'upstream'  spillovers? Analyzing this question in an open economy setting could provide interesting insights on how to foster long-run economic development.

\begin{appendix}
\numberwithin{equation}{section}
\appendix

   \section{Proof of Lemma \ref{lem:1}}\label{proof1}
By assumption, there is at least one $i\in\mathcal{I}$ with $\lim_{t\rightarrow\infty}F_{i}\:\mathbf{q}(t)+\alpha\rightarrow\infty$. If there are several such technologies, let us choose the technology $k$ for which $q_k(t)$ grows most rapidly in the long run, in the sense of the overtaking criterion. Dividing the nominator and denominator in Eq. \eqref{shares.1} by this $q_k(t)^{\frac{1}{1-\nu}}$ leads to  
\begin{equation}\label{shares.3}
	s_i(t)=\frac{\left(F_{i}\:\mathbf{q}_R (t)+\alpha/q_k(t)\right)^{\frac{1}{1-\nu}}}{\sum_{j=1}^N \left(F_{j}\:\mathbf{q}_R (t)+\alpha/q_k(t)\right)^{\frac{1}{1-\nu}}},
\end{equation}
where $\mathbf{q}_R (t):=\frac{1}{q_k (t)}\:\mathbf{q} (t)$. By construction, each component of $\mathbf{q}_R (t)$ converges to a value between $0$ and $1$ for $t\rightarrow\infty$. For all technologies that grow less than proportionally to $q_k(t)$ in the long run, this value is $0$. For all technologies $s$ that receive only spillovers from these 'slow growing'  technologies, the term $F_{s}\:\mathbf{q}_R (t)$ as well as the nominator in Eq. \eqref{shares.1} converges to zero. Thus, the share of scientists employed in developing these technologies converges to zero in the long run. These technologies form the set $\mathcal{I}_0$, which can be empty, if all technologies grow proportionally to $q_k(t)$.

The remaining technologies form the set $\mathcal{I}_\infty$. For these technologies, we denote the $i$-th component of the limit value of $\mathbf{q}_R (t)$ by $\zeta_{i}$ and observe that $\lim_{t\rightarrow\infty}\alpha/q_k(t)=0$. Using this information and taking the limit $t\rightarrow\infty$ in  Eq. \eqref{shares.3} directly yields Eq. \eqref{shares.2b}. Finally, observe that $\zeta_i$  denotes the long-run relative quality of technologies $i$ and $k$, where $k$ is the most rapidly growing technology. As the  $\zeta_i$ converge to constants in the long run, the growth rates of all technologies that continue to grow have to be identical.

   \section{Proof of Proposition \ref{prop.lrgrowth}}\label{proof1b}

	We start with Assertion 1.  Eqs. \eqref{eq.lr-2a}--\eqref{eq.lr-2b} always have the trivial solution $\mathbf{z}^{*}=0$. Assume that this would be the only solution. In this case, no technology would exhibit continued growth, which violates the assumption of Prop. \ref{prop.lrgrowth}. Thus, Assertion 1 has to hold.
	
	Lemma \ref{lem:1} implies that if two technologies exhibit continued growth, they have to grow at the same rate for $t\rightarrow\infty$. Thus for any two technologies $i,j$ for which this is the case, Eq. \eqref{eq.lr-1a} has to hold. If a technology stagnates, the corresponding $z_k$ equals zero, so that Eq. \eqref{eq.lr-1a}  holds as well. Given that Eq. \eqref{eq.lr-1b} is met by construction of the $z_i$, the system \eqref{eq.lr-1a}--\eqref{eq.lr-1b} is a necessary condition to identify technologies with continued growth. The system  \eqref{eq.lr-2a}--\eqref{eq.lr-2b} follows directly from this by calculating the growth rates from Eq. \eqref{qtoti} and Lemma \ref{lem:1} taking into account that the $z_i$ are proportional to the $\zeta_i$ in Lemma \ref{lem:1}. This proves Assertion 2.
	
	Assertion 3 is a consequence of Assertion 2 and Eqs. \eqref{eq.lr-2a}--\eqref{eq.lr-2b}. As there is only a single solution of Eqs. \eqref{eq.lr-2a}--\eqref{eq.lr-2b}, this solution fully characterizes the set of technologies that continue to grow according to Assertion 2. Furthermore, Eqs. \eqref{eq.lr-2a}--\eqref{eq.lr-2b} depend solely on the spillover structure, so that Assertion 3 has to hold.
	
	Finally, Assertion 4 follows from Assertion 2 and  Eqs. \eqref{qtoti} and \eqref{shares.1}. By Assertion 2, only solutions $\mathbf{z}^{*}\neq 0$ can characterize the set of technologies that continue to grow. If there are multiple candidates for such sets (i.e., different solutions $\mathbf{z}^{*}\neq 0$), Eqs. \eqref{qtoti} and \eqref{shares.1} determine which of these candidates actually characterizes long run growth. Apart from the spillovers, these equations depend only on the initial conditions and $\alpha$.

  \section{Proof of Proposition \ref{cor2}}\label{proof2}
  Assertion 1: If $\mathbf{F}=0$, the change of quality depends only on the number of scientists, not on qualities. Furthermore,  all firms are homogeneous and will thus each attract the same amount of scientists. As the number of scientists is constant, all firms will have the same strictly positive quality change in each period (i.e., $\alpha\:S/N$) and thus growth is linear and identical for all firms.

Assertion 2: 	If technologies are independent, then $\mathbf{F}$ is diagonal. In this case, Eq. \eqref{qtoti} for firm $i$ becomes independent of all qualities $q_j$, $j\neq i$. As we have at least one $F_{ii}>0$, at least one technology will grow exponentially. By Lemma \ref{lem:1}, only technologies with the highest value of $F_{ii}$ can attract scientists in the long run. These technologies will grow exponentially in the long run, all other technologies will stagnate.

Assertion 3: Under the conditions of Case (a), the matrix $\mathbf{F}$ can be brought into a lower triangular form. By the assumption $F_{ii}= 0$, the first technology can grow at most linearly in the long run, as it receives no spillovers and as, by  Lemma 1, the long-run shares of employed scientists are constant.  Thus the second technology, which receives at most a spillover from the first technology and for which $F_{ii}= 0$ also holds, can grow only polynomial. Repeating this argument for all technologies shows that growth can be at most polynomial. 

What remains to be shown is that growth is indeed restricted to linear growth. Suppose that some technologies would grow more than linearly  (e.g., quality being a quadratic function of time). Due to the linearity of the long-run dynamics and the structure of $\mathbf{F}$ in Case (a), this could only result from spillovers from technologies that (in the lower triangular form)  precede this technology. Let us assume that, in the lower triangular form, the first technology that shows more than exponential growth is technology $k$. Thus all technologies $i<k$ would grow slower in the long run, and thus, according to  Lemma \ref{lem:1}, their share of scientists goes down to zero. Thus they cease to provide a spillover to technology $k$ that enables more than linear growth, which implies that, in the long run, $k$ can only grow linearly. Repeating this argument for the next technology ($k+1$), and so on, shows that, in the long run, all technologies can only grow linearly.

 For Case (b) of Assertion 3, we have at least two technologies $i,j$ with $F_{ij}\:F_{ji}>0$. As spillovers are non-negative, these technologies grow in the least strongest way, if they receive no other spillovers. Furthermore, let us first assume that these are the most rapidly growing technologies and thus receive a constant share of scientists in the long run. Thus, the growth of these technologies is governed in the long run by a linear ODE system with a Jacobian that can be written as  \begin{equation}\begin{pmatrix}\label{eq.proofprop2}
0&F_{ij}\\
F_{ji}&0\\
\end{pmatrix}.
\end{equation}
This system always has a strictly positive eigenvalue ($\sqrt{F_{ij}\:F_{ji}}$) and thus shows exponential growth. By Lemma \ref{lem:1}, these two technologies would receive no scientists in the long run only if another technology grows more rapidly, so that we would have exponential growth for some technology. Finally, if there are additional spillovers, these technologies will grow even more rapidly. But as the model becomes linear in the long run, growth can be at most exponential. 

Assertion 4: If all technologies are strongly connected, each technology profits from a spillover from the most rapidly growing technology, implying that, in the long run, all technologies grow at the same rate. By Assertion 3 (b), this will be exponential growth. If spillovers  are homogeneous, then all firms attract the same number of scientists in the long run, so that Eq. \eqref{qtoti} implies that the growth rate of all qualities has to be proportional to $f$.

\end{appendix}

\end{document}